\documentclass[a4paper,physrev,twocolumn]{revtex4-2}
\usepackage{tikz}
\usepackage{amssymb}
\usepackage{amsmath}
\usepackage{bm}
\usepackage{braket}
\usepackage{graphicx}
\graphicspath{ {./figs/} }
\usepackage{color}
\usepackage{url}
\usepackage{siunitx}
\usepackage{hyperref}
\hypersetup{
  colorlinks=true,
    linkcolor=blue,
    filecolor=blue,      
    citecolor=blue,
    urlcolor=blue,
}
\newcommand{\I}{\mathrm{i}}
\newcommand{\E}{\mathrm{e}}
\newcommand{\D}{\mathrm{d}}

\DeclareMathOperator{\Tr}{tr}
\DeclareMathOperator{\CZ}{\mathsf{CZ}}
\DeclareMathOperator{\SW}{\mathsf{SW}}

\DeclareMathOperator{\C}{\mathsf{C}}
\newcommand{\T}{\mathsf{T}}
\DeclareMathAlphabet{\mathpzc}{OT1}{pzc}{m}{it}
\newcommand{\ee}{\mathpzc{e}}
\newcommand{\oo}{\mathpzc{o}}
\begin{document}

\title{Entanglement transition in a cluster spin chain coupled with free spins}
\author{Kevissen Sellapillay}\email{k.sellapillay@fz-juelich.de}
\affiliation{Jülich Supercomputing Centre, Institute for Advanced Simulation, Forschungszentrum Jülich, 52425, Jülich, Germany}
\author{Laurent Raymond}\email{laurent.raymond@univ-amu.fr}
\affiliation{Aix Marseille Univ, Université de Toulon, CNRS, CPT, Marseille, France}
\author{Alberto D.\ Verga}\email{alberto.verga@univ-amu.fr}
\affiliation{Aix Marseille Univ, Université de Toulon, CNRS, CPT, Marseille, France}

\date{\today}
\begin{abstract}
	We investigate the entanglement of a ladder of spins formed by two sublattices, a ``cluster'' chain and the ``environment'', consisting of independent spins, both coupled by an exchange interaction and evolving under a unitary discrete time dynamics. The automaton is defined by the composition of the two body spin swap gate (between sublattices) and the three body cluster interaction. We observe that, depending on the set of coupling constants, the cluster subsystem evolves towards states corresponding to different entanglement phases. In the weak coupling regime the subsystem remains near the topological cluster state. Increasing the coupling strength leads to random states which transform from almost pure to fully mixed, according to the effective number of the environment active degrees of freedom.
\end{abstract}
\maketitle

\section{Introduction}
\label{S:intro}

An important research trend at the interface of quantum information and condensed matter \cite{Zeng-2019} is the study of the entanglement structure of many-body states. Indeed, the entanglement of quantum phases reveals physical properties beyond their standard description in terms of order parameters and symmetry. For instance, the area law followed by the entanglement entropy found in gapped degenerated ground states, is characteristic of topological phases \cite{Kitaev-2003fk,Bravyi-2010,Wen-2017}. In contrast, for typical (non integrable) Hamiltonians, the entanglement entropy of highly excited states satisfies the volume law of thermal states. In fact, according to the eigenstate thermalization hypothesis \cite{Deutsch-1991vn,Srednicki-1994ys,Alessio-2016fj}, generic isolated quantum systems possess chaotic eigenstates, and naturally evolve to a thermal state. Remarkably, hybrid situations characterized by the breakdown of ergodicity, arise in some constrained systems such as low entropy states embedded in a thermal spectrum \cite{Bernien-2017,Turner-2018a,Papic-2022}, or fragmented Hilbert spaces \cite{Rakovszky-2020,Moudgalya-2022}, or can also arise dynamically in quantum cellular automata \cite{Sellapillay-2022b,Verga-2023}.

Topological states, especially those belonging to a symmetry protected topological phase, are useful resources for quantum computations. In the toric code \cite{Kitaev-2003fk} two qubits can be logically encoded using the degenerated ground state subspace; more generally, surface codes \cite{Bravyi-1998,Fowler-2012} with data and measurement qubits allowing active error-correction are suitable to the implementation of topological computing \cite{Freedman-2003,Lahtinen-2017fk}. Other important class of entangled topological states are the cluster states \cite{Briegel-2001fk,Raussendorf-2001uq}, which, using a convenient measurement protocol, transform into a resource for universal quantum computing \cite{Raussendorf-2012xr}.

However, on one side, random quantum states are not a useful resource for quantum computation \cite{Gross-2009uq}, and, on the other side, topological ground states not only are subject to decoherence at finite temperature \cite{Dennis-2002,Brown-2016}, but are difficult to prepare using local unitary circuits: long-range entanglement is limited by the ballistic spreading of entanglement (Lieb-Robinson limit \cite{Bravyi-2006, Hastings-2011}). A promising approach to generate useful quantum entangled states outside the domain of ground states or thermalized states is in the domain of out-of-equilibrium states. For instance, one may use stochastic quantum circuit protocols in which unitaries and measurements are randomly interspersed to reach a state with interesting entanglement structure \cite{Fisher-2023}.  When the rate of measurements exceeds some threshold a transition occurs from a highly entangled steady state to a low entangled one \cite{Li-2019,Gullans-2020,Choi-2020}. In fact, monitored random circuits exhibit a novel class of many-body states with distinct entanglement properties \cite{Bao-2021,Koh-2023,Tantivasadakarn-2024}.

Another interesting structure is observed in the entanglement of subsystems belonging to a global random states \cite{Lubkin-1978,Page-1993nr}, for both pure and mixed states \cite{Kendon-2002,Znidaric-2007}, namely, the partial transposed density matrix (negativity) spectrum of the subsystem transits from a Marčenko-Pastur to a Wigner distribution depending on its size. A related effect displays in bipartite systems in a random state with fixed von Neumann entropy: the density matrix (entanglement) spectrum acquires the shape of a deformed Marčenko-Pastur distribution according to their entanglement level \cite{Facchi-2013}. More specifically, the negativity spectrum of the subsystem density matrix, obtained from a tri-partition of a pure random state, shows a transition from an entangled phase for small subsystem size, to an unentangled one above a threshold in the subsystem size \cite{Bhosale-2012,Shapourian-2021}. This transition was recently simulated in a quantum computer using all-to-all swap gates over 15 qubits \cite{Liu-2023}, reminiscent to the entanglement transition of monitored circuits. We then realize that even seemingly unstructured, random states, may exhibit non-trivial entanglement patterns depending, for instance, on the number of degrees of freedom of the surrounding \cite{Znidaric-2007,Aubrun-2012}.

In this paper we explore the entanglement transitions of a cluster spin chain \cite{Raussendorf-2005} interacting with free spins mimicking an environment, the two subsystem forming a spin ladder (see Fig.~\ref{f:lat}). Instead of continuously monitoring the environment of a unitary random circuit, we tune the entanglement properties of the cluster subsystem by varying its coupling with the free spins. The whole system unitarily evolves following a discrete time dynamics (Sec.~\ref{S:model}), and can be represented by an automaton with a two sites elementary cell (c.f.\ Fig.~\ref{f:automat}). We show that this simple model can describe the transition between the cluster topological phase and a high energy thermal phase, and also it can account for transitions between different random states, depending on the initial state and the strength of the coupling between the two subsystems (see the results of Sec.~\ref{S:results}).

\section{Model}
\label{S:model}

\begin{figure}[tbp]
  \centering%
  \includegraphics[width=0.49\textwidth]{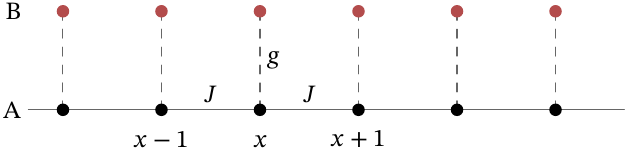}
  \caption{Bipartite lattice, each dot represents a spin 1/2 (qubit). The A sites (lower row) interact via a three-body interaction of energy $J$, the cluster interaction; the B sites (higher row) are free; the two sublattices, cluster and environment, are coupled by the exchange interaction of energy $g$.}
  \label{f:lat}
\end{figure}

We consider a bipartite one dimensional lattice of $L$ cells $x=1,2, \cdots, L$; each cell site is occupied by a one-half spin, forming two sublattices that are denoted A and B. We take the lattice step as the length unit. Sublattice A is identified as the ``cluster'' subsystem, or, depending on the context, simply the ``system'', and sublattice B as the ``environment'' (Fig.~\ref{f:lat} represents the lattice geometry and notation). We mostly assume periodic boundary conditions.

\begin{figure}[tbp]
  \centering%
  \includegraphics[width=0.49\textwidth]{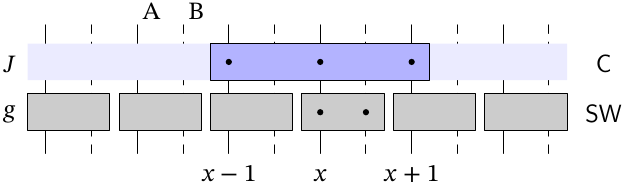}
  \caption{One time step of the automaton: $J$ is the characteristic energy of the cluster sublattice A, the support of the $\C_\textsc{a}$ operator; $g$ couples the two sublattices exchanging and superposing their states through the $\SW_\textsc{ab}$ interaction. Note that the $C_\textsc{a}$ operators commute, therefore their are applied on all A sites simultaneously.}
  \label{f:automat}
\end{figure}

The Hilbert space of the system has the structure of an interleaved tensor product between the spin spaces of the two sublattices,
\begin{equation}
  \label{e:hilbert}
  \mathcal{H} = \mathcal{H}_\textsc{a} \otimes \mathcal{H}_\textsc{b} = \bigotimes_{x=1}^L \mathcal{H}_\textsc{a}^{(x)} \otimes \mathcal{H}_\textsc{b}^{(x)}, 
\end{equation}
where the second equality corresponds to the AB cell decomposition of the Hilbert space. We use the notation $(X,Y,Z)$ for the standard Pauli matrices; for instance, $X_x^\textsc{a}$ acts on the total Hilbert space $\mathcal{H}$ as the identity on all sites but the A site of cell $x$, where $X$ is applied, etc. The dynamics of the system is discrete in time, local, and translation invariant, it is defined by the unitary automaton one time step operator $U$ (Fig.~\ref{f:automat}),
\begin{equation}
  \label{e:U}
  U(J,g) = \prod_{x=1}^L \C_\textsc{a}(J, x) \prod_{x=1}^L \SW_\textsc{ab}(g, x),
\end{equation}
($\Delta t =1$ and $\hbar=1$) where
\begin{equation}
  \label{e:CA}
  \C_\textsc{a}(J, x) = \exp\big(\I J Z_{x-1}^\textsc{a} X_x^\textsc{a} Z_{x+1}^\textsc{a}\big)
\end{equation}
is the cluster interaction operator and $J$ the three-body coupling energy, formed by the complete set of stabilizers of the form $ZXZ$ \cite{Hayden-2006},
and
\begin{equation}
  \label{e:SW}
  \SW_\textsc{ab}(g, x) = \exp\big[\I g (X_x^\textsc{a} X_x^\textsc{b} + Y_x^\textsc{a}Y_x^\textsc{b})/2\big],
\end{equation}
is the swap operator which exchange the spins between the two sublattices ($g$ is the exchange coupling energy).

It is important to remark that the distinction between ``cluster'' and ``environment'' is justified by the difference in the sublattices interactions; however, the free spin sublattice, at variance to a standard environment of an open system, is here completely determined by its interaction with the cluster chain and, in particular, may act coherently (see below the discussion about the Markov approximation of the automaton).

In the following we investigate the properties of the automaton \eqref{e:CA} using two complementary approximations. In the low energy limit, we use a mean-field approximation that captures the entanglement transition due to the perturbation of the topological phase of the cluster subsytem. At higher energies, we assume that the system is in a random phase, and therefore approximate its dynamics using a Markov approximation. It is worth stressing that while the mean-field is essentially perturbative, the Markov master equation is not limited to small values of the coupling constant $g$: its validity depends on the assumption of short temporal correlations (the state of the cluster subsystem depends only on the last state of the system).

We start by studying the continuous time limit of the automaton. In this limit, when $U$ is near the identity (formally $J,g \rightarrow 0$), the system is described by the time independent Hamiltonian
\begin{multline}
  \label{e:H}
  H = H_\mathsf{C} + H_\mathsf{SW} = -J\sum_{x = 1}^L Z_{x-1}^\textsc{a} X_x^\textsc{a} Z_{x+1}^\textsc{a} -\\
   \frac{g}{2} \sum_{x = 1}^L  \big(X_x^\textsc{a} X_x^\textsc{b} + Y_x^\textsc{a} Y_x^\textsc{b}\big),
\end{multline}
where the first term $H_\mathsf{C}$ is the cluster Hamiltonian \cite{Pachos-2004,Raussendorf-2005,Skrovseth-2009,Smacchia-2011,Montes-2012}, a sum of stabilizers of the cluster state
\[
\ket{C_A} = \prod_{x=1}^L \CZ_\textsc{a}(x,x+1 \mod L) \ket{+}^{\otimes L},
\]
which is then its ground state: here $\CZ = \mathrm{diag}(1,1,1,-1)$ is the two qubits phase gate and $\ket{+}$ the 1 eigenstate of $X$; the second term $H_\mathsf{SW}$ describes the coupling of $\textsc{A}$ and $\textsc{B}$ sublattices. This second term, because of the presence of $Y_x^\textsc{a}$, breaks the $\mathbb{Z}_2\times\mathbb{Z}_2$ symmetry of the cluster subsystem \cite{Smacchia-2011}. However, the whole system possesses a $\mathbb{Z}_2$ symmetry generated by
\begin{equation}
  \label{e:PAB}
  P_{\textsc{ab}} = \prod_{x=1}^L X_x^\textsc{a}X_x^\textsc{b}, \quad P_\textsc{ab}UP_\textsc{ab} = U,
\end{equation}
which obviously translates to $H$. Note that there is also a spectral mirror symmetry
\begin{equation}
  \label{e:symA}
  C_{\textsc{a}} = \prod_{x=1}^L Z_x^\textsc{a}, \quad C_\textsc{a}UC_\textsc{a} = U^\dagger,
\end{equation}
which changes $H \rightarrow -H$ of Eq.~\eqref{e:H}, or equivalently $U \rightarrow U^\dagger$.

Using the Jordan-Wigner transformation to write the Pauli operators in terms of $a_x$ and $b_x$, the annihilation operators of a fermion on sites A and B of cell $x$, respectively,
\begin{align}
  \label{e:JW}
  &X_x^\textsc{a} = 1 - 2 a^\dagger_x a_x = s_x^\textsc{a}, &  X_x^\textsc{b} = 1 - 2 b^\dagger_x b_x = s_x^\textsc{b}, \nonumber\\
  &Y_x^\textsc{a} = \I N_x^\textsc{a} (a^\dagger_x - a_x), & Y_x^\textsc{b} = \I N_x^\textsc{b} (b^\dagger_x - b_x), \\
  &Z_x^\textsc{a} = - N_x^\textsc{a} (a^\dagger_x + a_x), & Z_x^\textsc{b} = - N_x^\textsc{b} (b^\dagger_x + b_x), \nonumber
\end{align}
with 
\[
N_x^\textsc{a} = \prod_{y = 1}^{x-1} \big(1 - 2a^\dagger_y a_y \big) \big(1 - 2b^\dagger_y b_y \big)= \prod_{y = 1}^{x-1} s_y^\textsc{a} s_y^\textsc{b}
\]
($s_x$ has eigenvalues $\pm 1$)
and
\[
N_x^\textsc{b} = N_{x}^\textsc{b}(1 - 2a^\dagger_{x} a_{x}) = N_{x}^\textsc{a} s^\textsc{a}_{x},
\]
we obtain the local fermionic Hamiltonian,
\begin{multline}
  \label{e:Hffff}
  H = -\sum_{x = 1}^L \Big\{J (a^\dagger_{x-1} - a_{x-1}) s_{x-1}^\textsc{b} s_x^\textsc{b} (a^\dagger_{x+1} + a_{x+1}) \\
  + g [s_x^\textsc{a} s_x^\textsc{b} - (a^\dagger_x + a_x) (b^\dagger_x - b_x)] \Big\}.
\end{multline}
It is worth noting that the present model, even in this limit, is qualitatively different to the previously studied variants of the cluster model, in that the cluster Hamiltonian contains a dynamical interaction with the environment through its spin field $s_x^\textsc{b}$.

\subsection{Mean field approximation}

The low energy properties of $H$ can be studied in the mean-field approximation. The natural mean-field order parameters is the $H$ ground state expectation value $s_\textsc{b}=\braket{s_x^\textsc{b}}$ of the environment spin:
\begin{align}
  \label{e:HMF}
  H_\mathsf{MF} &= -\sum_{x = 1}^L \Big[ Js_\textsc{b}^2 (a^\dagger_{x-1} - a_{x-1}) (a^\dagger_{x+1} + a_{x+1}) + g s_\textsc{b} s^\textsc{a}_x \Big] \nonumber \\
  & \qquad {}  + g \sum_{x = 1}^L (a^\dagger_x + a_x) (b^\dagger_x - b_x) \nonumber\\
    &= H_\mathsf{C} + H_\mathsf{SW} 
\end{align}
The first sum ($H_C$) is analogous to a $Js_\textsc{b}^2$ cluster model in a $x$-field $gs_\textsc{b}$. The second sum ($H_\mathsf{SW}$) corresponds to a magnetic coupling with the environment B \cite{Pachos-2004}. Note that the $g$ coupling splits into two terms, one (originating from the $X^\textsc{a} X^\textsc{b}$ term) contributes to an effective applied field, and the other (arising from $\sim Y^\textsc{a} Y^\textsc{b}$) dephases and flips the A spins \cite{Smacchia-2011}. The phase diagram of various related cluster spin models is known (e.g.\ Ref.~\cite{Montes-2012}).

The parity symmetry of the original spin Hamiltonian translates to the fermionic one into
\begin{equation}
  [H,N_L] = 0, \quad N_L = N_L^\textsc{a},
\end{equation}
where $N_L$ is the parity operator (c.f.\ Eq.~\eqref{e:JW}). Therefore, the Hilbert space splits into two sectors, even ($N_L=1$) and odd ($N_L=-1$), according to the total number of fermions. Translational invariance allows us to perform a Fourier transform of the fermion operators:
\begin{equation}
  \label{e:Fou}
  a_x = \frac{\E^{\I \pi/4}}{\sqrt{L}} \sum_k \E^{\I k x} a_k,
\end{equation}
where $k \in \ee$ in the even sector, and $k \in \oo$ in the odd one,
\begin{gather}
\ee = \Big\{k = \frac{2\pi}{L}(n - 1/2) \mid n = -L/2+1,\ldots, L/2\Big\} \\
\oo = \Big\{k = \frac{2\pi}{L}n \mid n = -L/2+1,\ldots, L/2\Big\} .
\end{gather}
(We choose an even number of cells $L$.) This leads to the mean-field Hamiltonian in Fourier space
\begin{multline}
  \label{e:HkC}
  H_\mathsf{C} = -2 \sum_{k>0} \Big[ \big(Js_\textsc{b}^2 \cos2k - gs_\textsc{b} \big) \big(a_k^\dagger a_k - a_{-k} a_{-k}^\dagger\big) \\
  + Js_\textsc{b}^2 \sin2k \big(a_k^\dagger a_{-k}^\dagger - a_k a_{-k}\big)\Big],
\end{multline}
where the sum runs over positive $k$, and 
\begin{equation}
  \label{e:HkSW}
  H_\mathsf{SW} = -g\sum_{k>0} \Big(\I a\strut_{(k}^\dagger b\strut_{-k)}^\dagger + \I a\strut_{(k} b\strut_{-k)} + a\strut_{(k}^\dagger b\strut_{k)} - a\strut_{(k} b\strut_{k)}^\dagger\Big),
\end{equation}
where the parentheses mean symmetrization $k \leftrightarrow -k$. Introducing the spinor 
\[
  C_k = \begin{pmatrix}
    a_k \\ a_{-k}^\dagger \\ b_k \\ b_{-k}^\dagger 
  \end{pmatrix}
\]
the mean field Hamiltonian can be written in the matrix form
\begin{equation}
  \label{e:HkMF}
  H_{\mathsf{MF}} = \sum_{k>0} C_k^\dagger \big[H_\mathsf{C}(k) + H_\mathsf{SW}(k)\big] C_k
\end{equation}
where 
\begin{multline}
  \label{e:HkCM}
  H_\mathsf{C}(k) = -2\big(J s_\textsc{b}^2\cos 2k - g s_\textsc{b}) \frac{1+Z}{2} \otimes Z \\
  - 2J s_\textsc{b}^2\sin 2k \frac{1+Z}{2}  \otimes X,
\end{multline}
and
\begin{equation}
  \label{e:HkSWM}
  H_\mathsf{SW}(k) = g Y \otimes X - g X \otimes Z.
\end{equation}

\begin{figure*}[t]
  \includegraphics[width=0.48\textwidth]{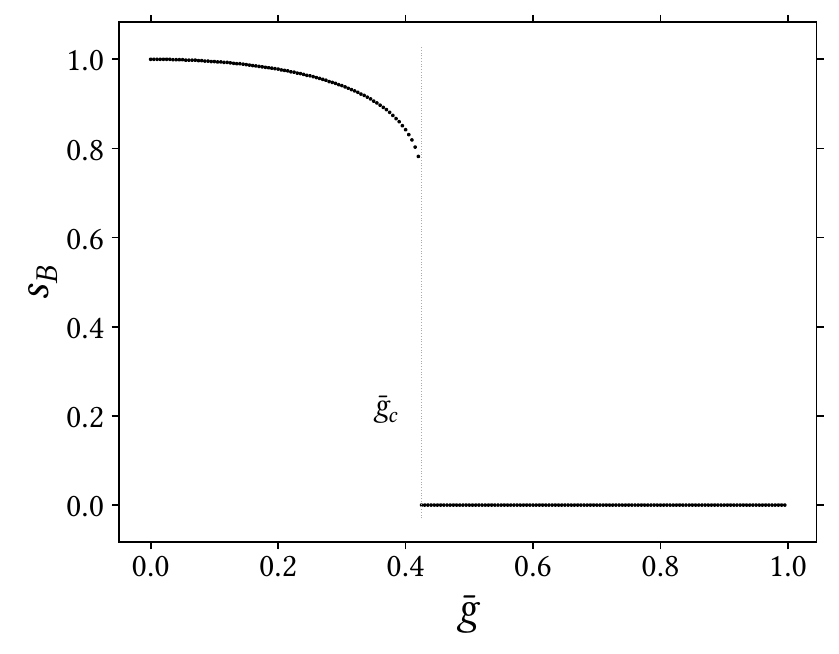}\includegraphics[width=0.48\textwidth]{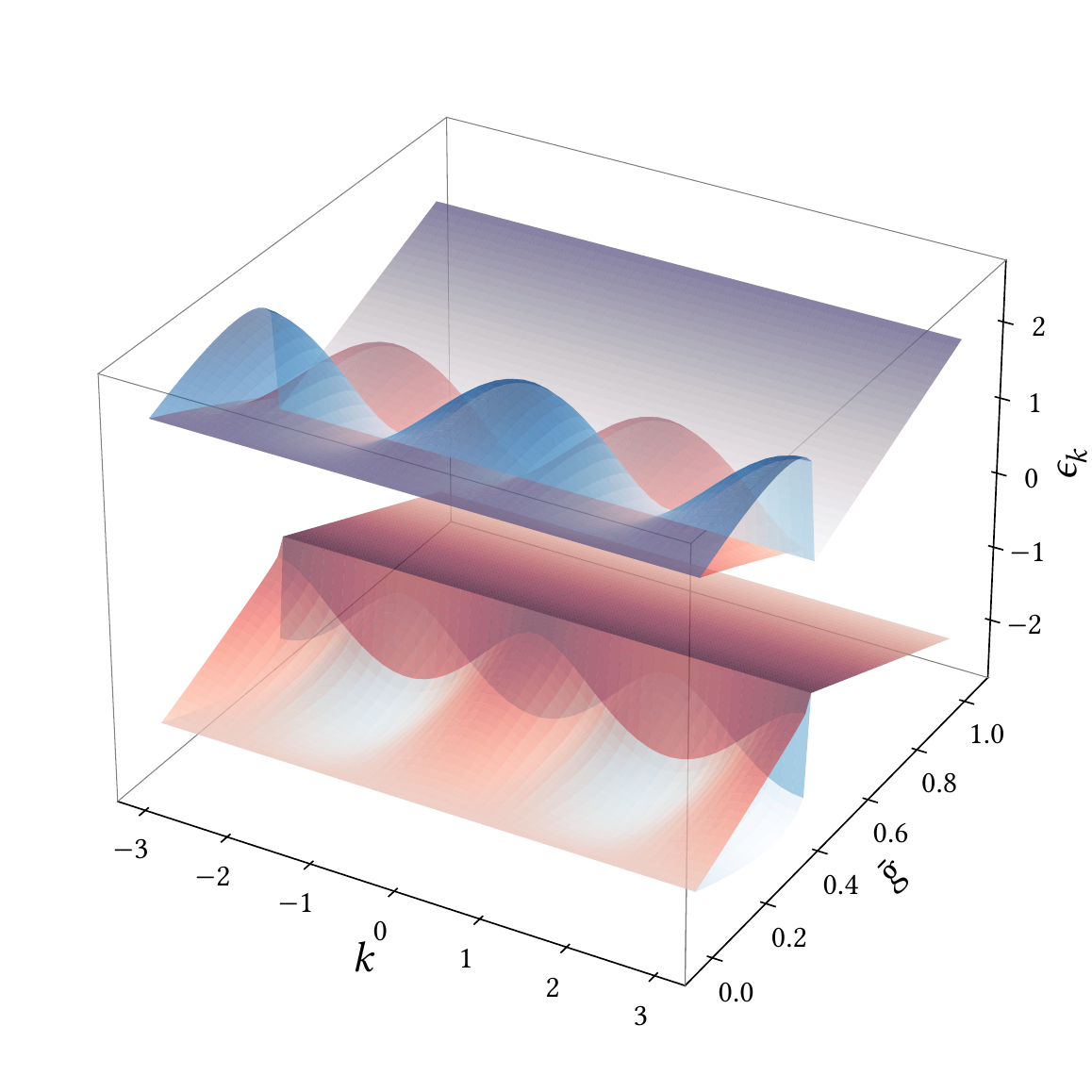}
  \caption{Band structure in the mean field approximation. (left) Mean field $s_\textsc{b}$ \protect{\eqref{e:sb}} as a function of $\bar{g}=g/J$. (right) Dispersion relation $\epsilon_k(\bar{g})$ \protect\eqref{e:ek}. Blue and red interleaving surfaces correspond to the two sings of $s_\textsc{b}$. For $\bar{g} > 0.425$, beyond the discontinuity of $\epsilon_k(\bar{g})$, the bands become flat.\label{f:ek}}
\end{figure*}

The mean field energy spectrum can be computed from the full matrix,
\begin{widetext}
\begin{equation}
  \label{e:HkMFm}
  H_\mathsf{MF}(k) = \begin{pmatrix}
   -2Js_\textsc{b}^2 \cos 2k + g s_\textsc{b} & -2Js_\textsc{b}^2 \sin 2k & -g & - \I g \\
   -2Js_\textsc{b}^2 \sin 2k & 2Js_\textsc{b}^2 \cos 2k - g s_\textsc{b} & -\I g & g \\
   -g & \I g & 0 & 0 \\
   \I g & g & 0 & 0 
  \end{pmatrix}
\end{equation}
\end{widetext}
by diagonalizing \eqref{e:HkMF} using a standard Bogoliubov transformation $\T$ \cite{Franchini-2017} of the fermion operators
\begin{equation}
  \label{e:T}
   \begin{pmatrix}
    A_k \\ A_{-k}^\dagger \\ B_k \\ B_{-k}^\dagger 
  \end{pmatrix} = \T^\dagger
   \begin{pmatrix}
    a_k \\ a_{-k}^\dagger \\ b_k \\ b_{-k}^\dagger 
  \end{pmatrix},
\end{equation}
such that $\T^\dagger H_\mathsf{MF}\T$ is diagonal.

We obtain two 0 eigenvalues and two symmetric eigenvalues $\pm \varepsilon_k$, with dispersion
\begin{equation}
  \label{e:ek}
  \varepsilon_k = 2 \sqrt{(J s_\textsc{b} - g)^2 s_\textsc{b}^2+ g^2 + 4Js_\textsc{b}^3 g \sin^2 k},
\end{equation}
which is even in $k$, and then can be easily extended to the full Brillouin zone. Dispersion relations of this type were associated to the existence of continuous quantum phase transition, without gap closing \cite{Ezawa-2013}; however, in our case the presence of the self-consistent environment magnetization as an extra parameter, modifies this picture as we will show.

To compute explicitly the dispersion relation we need to find $s_\textsc{b}$; this is donne in Appendix~\ref{S:app_sb}. We obtain the self-consistent equation,
\begin{equation}
  \label{e:sbint}
  s_\textsc{b} = \int_{-\pi}^\pi \frac{\D k}{2\pi} \frac{2g^2}{\varepsilon_k^2 - 2g^2} - 1,
\end{equation}
which, after computing the elementary integral (note that $\{g \rightarrow -g, s_\textsc{b} \rightarrow -s_\textsc{b}\}$ lets the integral invariant), leads to the algebraic equation
\begin{multline}
  \label{e:sb}
  s_\textsc{b} = \pm 1 \mp \\
    \frac{\bar{g}^2}{\sqrt{4 s_\textsc{b}^8 - 8 \bar{g}^2 s_\textsc{b}^6 + 4 \bar{g}^2 (1 + \bar{g}^2) s_\textsc{b}^4 + \bar{g}^4 (4 s_\textsc{b}^2 + 1)}}
\end{multline}
where the two signs correspond to negative and positive $s_\textsc{b}$, and $\bar{g}=g/J$ is the only free parameter; $s_\textsc{b} = 0$ is always a solution, and, depending on $\bar{g}$ there are two pairs of opposite sign roots. In particular, for $g=0$ we find $s_\textsc{b} = \pm 1$, and the mean-field Hamiltonian reduces to the cluster one, $H_\mathsf{MF} = H_\mathsf{C}$. At $g = 0.425 \, J = \bar{g}_c J$, the upper branch ends at the value $s_\textsc{b}=0.73$, and the $s_\textsc{b}=0$ root branch starts and becomes the stable one, signaling a first order phase transition, as shown in Fig.~\ref{f:ek} (left).

In Fig.~\ref{f:ek} (right), we show the dispersion relation \eqref{e:ek} obtained by solving the self-consistent equation \eqref{e:sb}. We observe that for $\bar{g}>0.425$, the two bands become flat, independent of the wavenumber and proportional to the coupling with the environment, $\epsilon_k \sim \bar{g}$. In the small $\bar{g}$ region, we find two pairs of interleaved bands, separated by a gap which reach its maximum value 2 for $g=0$ and its minimum $2\bar{g}_c$ at the critical point.  These two qualitatively different sectors of the energy spectrum, correspond to a spin polarized (for $\bar{g} < \bar{g}_c$) and unpolarized (for $\bar{g} > \bar{g_c}$) environment. Therefore, in the mean-field approximation, $\bar{g}_c$ is the critical parameter of a first-order quantum phase transition, through which the topology of the ground-state, and the magnetic properties of the system change. In the following sections we analyse how these properties influence the system's \emph{dynamics} and in particular its long-time stationary states.

\subsection{Markov approximation}

In order to go beyond the properties of the system in the ground state, we investigate its dynamics using a quantum channel Markov approximation. We thus trace out the environment degrees of freedom, those associated with the B free spins, to obtain the change of the system's density matrix after one step $\Delta t = 1$,
\begin{equation}
  \label{e:markov} 
  \rho_\textsc{a}(t + 1) = \Tr_\textsc{b} \big[ U\rho_\textsc{ab}(t) U^{\dagger}\big],
\end{equation}
with $U$ the automaton unitary operator \eqref{e:U} and $\rho_\textsc{ab}$ the (pure) system's state. This can be written in terms of Kraus $M_n$ operators \cite{Schlosshauer-2019,Pearle-2012kl,Gillman-2023}
\begin{equation}
  \label{e:markovMn}
  \rho_\textsc{a}(t + 1) = \sum_n M_n \rho_\textsc{a}(t) M_n^\dagger, \quad
    M_n = \braket{n_B|U|t_B}
\end{equation}
in the Markovian approximation, where $\ket{t_B}$ is the environment state at time $t$ and $\ket{n_B}$ is a basis state of the environment (in the $\ket{\pm}$ basis at each site). The basic approximation is that at each time step the density matrix $\rho_\textsc{ab} = \rho_\textsc{a} \otimes \rho_\textsc{b}$, can be written as a product of the system state $\rho_\textsc{a}(t)$ and the environment state $\rho_\textsc{b}(t)$. Only the swap operators $\SW_\textsc{ab}(g, x)$ are affected by the expectation value in \eqref{e:markovMn}, given at each site
\begin{equation}
  \label{e:sw0}
  \bra{+_\textsc{b}}\SW_\textsc{ab}(g,x)\ket{+_\textsc{b}} = \cos(g/2)  e^{igX^\textsc{a}_x/2}
\end{equation}
and
\begin{equation}
  \label{e:sw1}
  \bra{-_\textsc{b}}\SW_\textsc{ab}(g,x)\ket{+_\textsc{b}} = \sin(g/2) e^{-igX^\textsc{a}_x/2} Y^\textsc{a}_x,
\end{equation}
where we used the identity
\[
  \SW(g) = \cos^2 \frac{g}{2} + \sin^2 \frac{g}{2} ZZ + \I \sin g (XX + YY).
\]
Therefore, the Kraus operators can be written as (we drop the unnecessary index A of the cluster sublattice operators),
\begin{equation}
  \label{e:Mn}
  M_n = u_0(g) \cos^{L-|n_+|} \frac{g}{2} \sin^{|n_+|} \frac{g}{2} \Big(\prod_{x\in n_+} \E^{-\I gX_x} Y_x \Big)
\end{equation}
where we defined $n_+ = \{x \mid n_{x-1}=1\}$, the set of 1 positions in the binary expansion
\[
n = \sum_{x=1}^L n_{x-1} 2^{L - x}, \quad n = 0, \ldots, 2^L - 1,
\]
and $|n_x|=0,1,\ldots,L$ the numbers of $1$ in this expansion, corresponding to the number of flips $\ket{+} \rightarrow \ket{-}$ associated with the state $\ket{n_\textsc{b}}$. Note that $M_n$ contains the operator
\begin{equation}
  u_0(g) = \prod_x \exp\Big[ \I J Z_{x-1} X_x Z_{x+1} + \frac{\I g}{2} X_x \Big],
\end{equation}
which after replacement in the Markovian equation (see Eq.~\eqref{e:markovMnp} below), appears as an effective unitary evolution operator in the presence of the environment; it defines the effective Hamiltonian
\begin{equation}
  \label{e:lindH0}
  H_0 = -J \sum_x Z_{x-1} X_x Z_{x+1} - \frac{g}{2} \sum_x X_x,
\end{equation}
of a cluster system in a $g/2$ applied field. We verify that
\begin{multline*}
  \sum_n M^\dagger_n M_n = \\
  \sum_{|n_x|=0}^{L} \binom{L}{|n_+|} \big(\cos^2\frac{g}{2}\big)^{L-|n_+|} \big(\sin^2\frac{g}{2}\big)^{|n_+|} = 1.
\end{multline*}

\begin{figure*}[t]
	\centering%
  \includegraphics[width=0.48\textwidth]{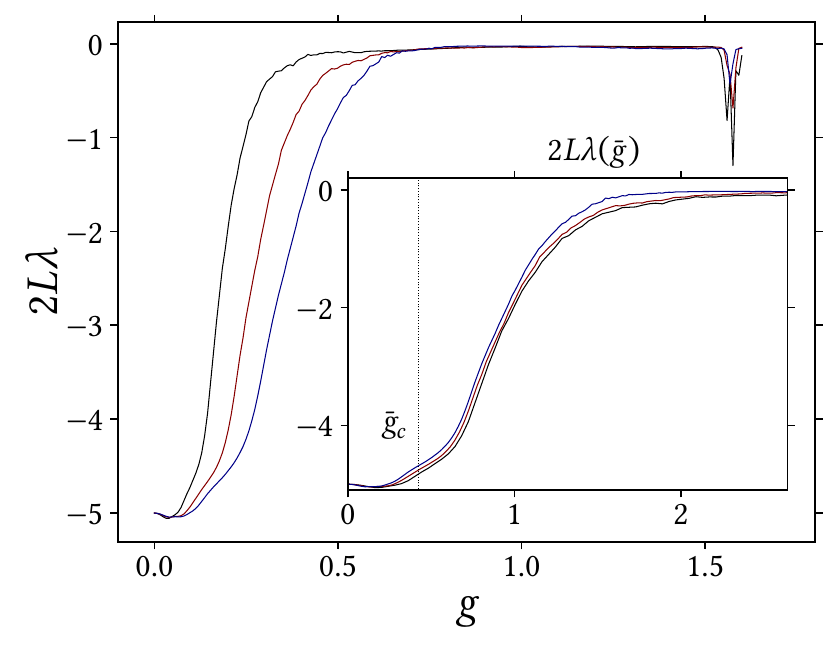} \includegraphics[width=0.48\textwidth]{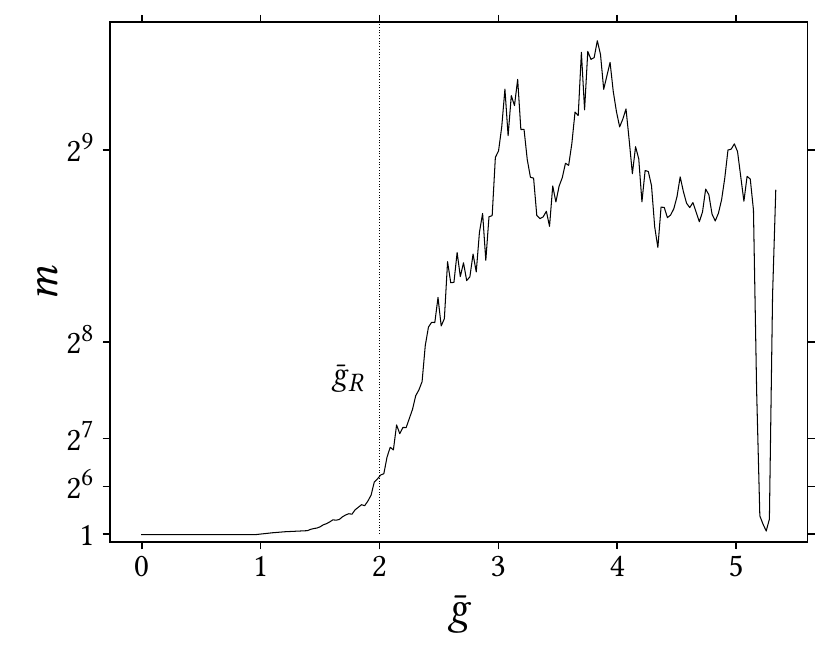}
	\caption{Entanglement witness and environment coupling driven random states. (left) Minimum negative eigenvalue $\lambda$ as a function of the environment coupling constant $g$, for three values $J=0.2,0.3,0.4$, from left to right ($2L=20$). The inset shows the scaling with $\bar{g}=g/J$, and the mean-field critical parameter $\bar{g}_c$. (right) Number of the environment effective degrees of freedom interacting with the cluster subsystem; above $\bar{g}_R$ the cluster subsystem is in a random, weakly entangled, mixed state, according to the Markov approximation ($J=0.3$).}
	\label{f:lg}
\end{figure*}

Inserting the expression of $M_n$ \eqref{e:Mn} into the Markov equation \eqref{e:markovMn} we get
\begin{widetext}
\begin{equation}
  \label{e:markovMnp}
  \rho_\textsc{a}(t + 1) = u_0(g) \Big[ \sum_{|n_x|=0}^{L} \binom{L}{|n_+|} \big(\cos^2\frac{g}{2}\big)^{L-|n_+|} \big(\sin^2\frac{g}{2}\big)^{|n_+|} \Big(\prod_{x\in n_+} \E^{-\I gX_x} Y_x \Big) \rho_\textsc{a}(t) \Big(\prod_{x\in n_+} Y_x \E^{\I gX_x} \Big) \Big] u_0^\dagger(g).
\end{equation}
\end{widetext}
We note that the interaction with the environment, in the Markovian limit, modifies the unitary evolution of the cluster system by the action of a $g/2$ field, suggesting the existence of a dynamical transition towards a random phase at $\bar{g}=2=\bar{g}_R$. The environment also tends to decohere the cluster subsystem state through irreversible phase damping: each term in \eqref{e:markovMnp} corresponds to a sequence of ``errors'' ($Y$-flips).

Usually, in the Lindblad approximation, one considers stochastic processes in which the probability of simultaneous errors are negligible (c.f.\ Appendix~\ref{S:app_lind}). At variance, for initially homogeneous states, translation invariance implies that the relevant term is instead the one in which we have $L$ simultaneous $Y$ flips of the environment spins, this ensures perfect synchronization of the homogeneous environment state. In this case the Markov equation \eqref{e:markovMnp} simplifies to
\begin{equation}
\label{e:ML}
  \rho(t+1) = u_0(-g) \sin^{2L}\frac{g}{2} \Big(\prod_x Y_x \Big) \rho(t) \Big(\prod_x Y_x \Big) u_0^\dagger(-g).
\end{equation}
We remark that, in this equation, the coherent evolution $u_0(-g)$, introduces the same field $\bar{g}/2$ present in ($\ref{e:Mn}$) but with the opposite sign. Therefore, the Markov equation ($\ref{e:ML}$) not only breaks the mean-field symmetry $g \rightarrow -g$, but the effective field tends to polarize the cluster spins in the $\ket{-}$ eigenstate.

To appreciate the impact of the independent spins on the properties of the cluster states it is important to identify the symmetries of the Markov equations \eqref{e:markovMnp}-\eqref{e:ML}. Indeed, while errors satisfying strong symmetries preserve the topology of the cluster phase, weak symmetry breaking jumps will eventually destroy the topological phase. Strong symmetries are associated to jump operators that commute with both the cluster Hamiltonian and its symmetry operators and weak symmetries do not imply a relationship with conserved quantities \cite{Buca-2012,Paszko-2024}. In our case we have 
\begin{equation}
  \label{e:PMP}
  [H_0,P] = 0, \; P M_n P=(-1)^{|n_+|} M_n, \quad P = \prod_x X_x,
\end{equation}
jumps with $n_+$ even are strong symmetries of the master equation, while they are weak symmetries for $n_+$ odd. As a consequence, a coherent phase flip $M_L$ of all spins does not break the cluster parity symmetry (the number of spins is even), extending the symmetry protected topological phase beyond the regime of vanishing coupling with the environment. This kind of ``coherent'' errors is relevant for initial homogeneous states, homogeneity that is preserved by the translational invariant dynamics. In the extreme Lindblad limit, incoherent localized jumps tends to relax the subsystem towards a vanishing entanglement state (see Appendix~\ref{S:app_lind}). In the next section we numerically investigate the behavior of the cluster subsystem by varying the interaction strength with the free spins of the environment.

\begin{figure*}[tbp]
  \centering%
  \includegraphics[width=0.3\textwidth]{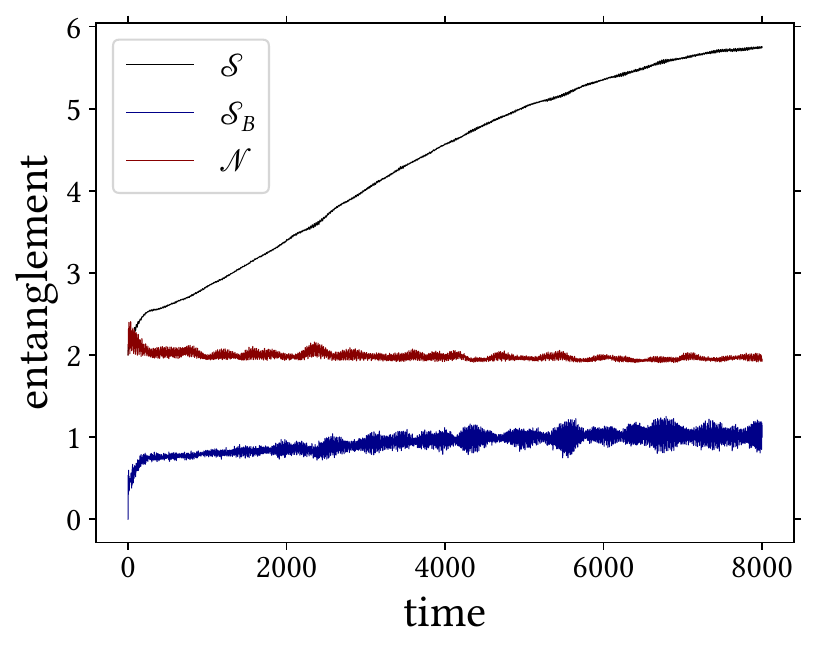}%
  \includegraphics[width=0.3\textwidth]{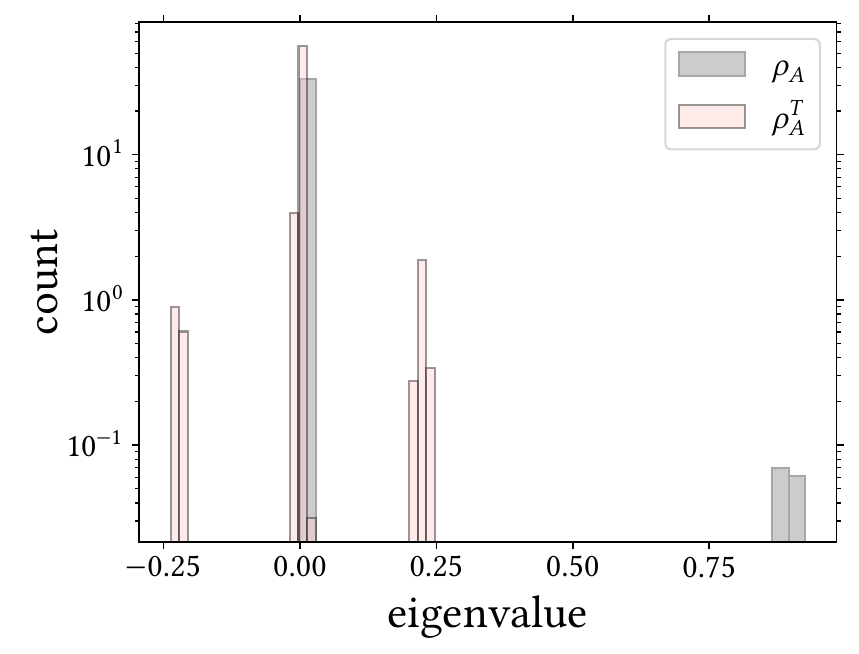}%
  \includegraphics[width=0.3\textwidth]{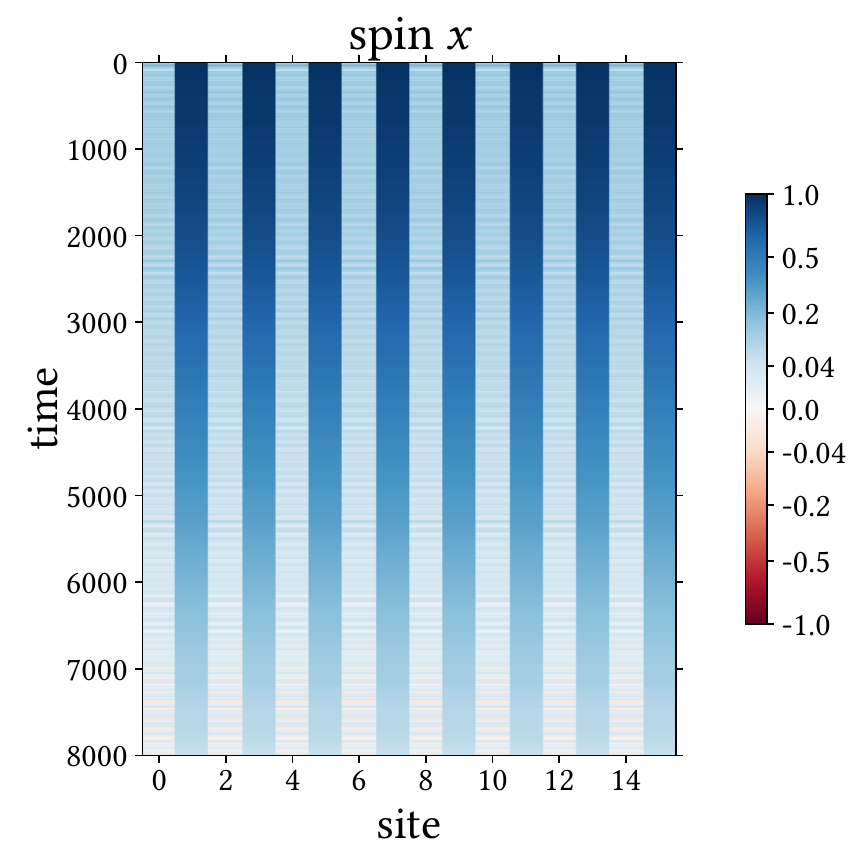}\\
  \includegraphics[width=0.3\textwidth]{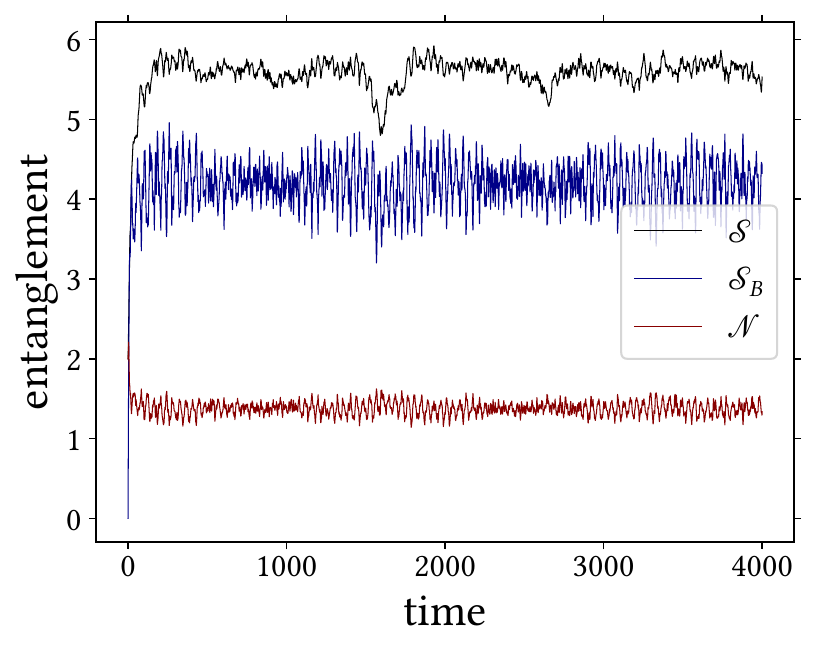}%
  \includegraphics[width=0.3\textwidth]{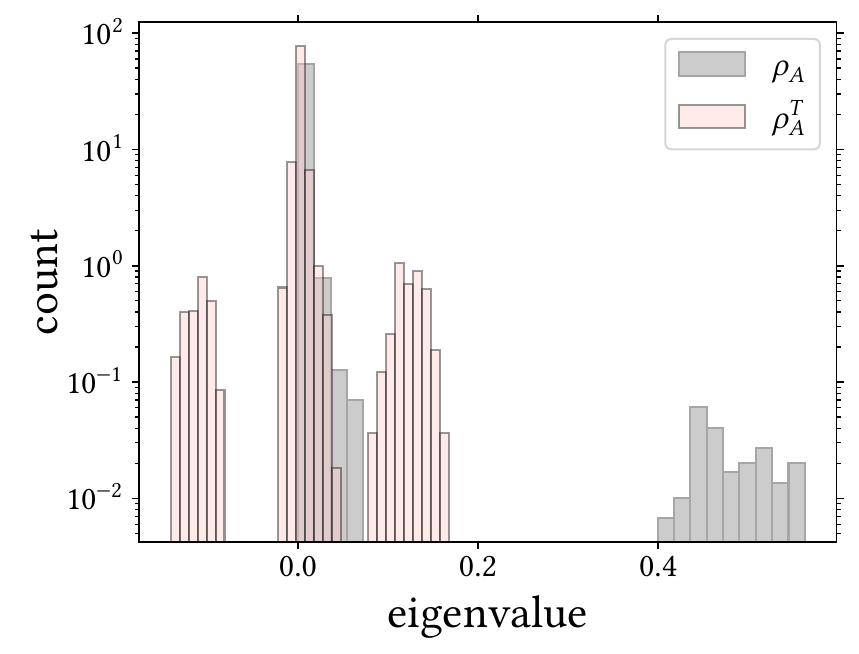}%
  \includegraphics[width=0.3\textwidth]{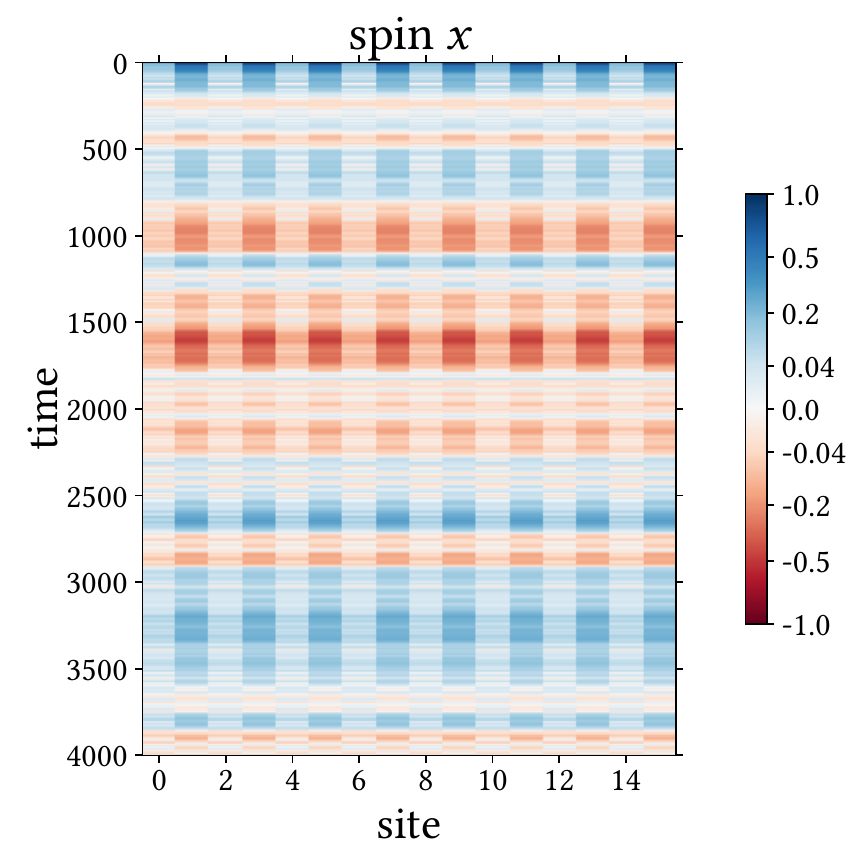}\\
  \includegraphics[width=0.3\textwidth]{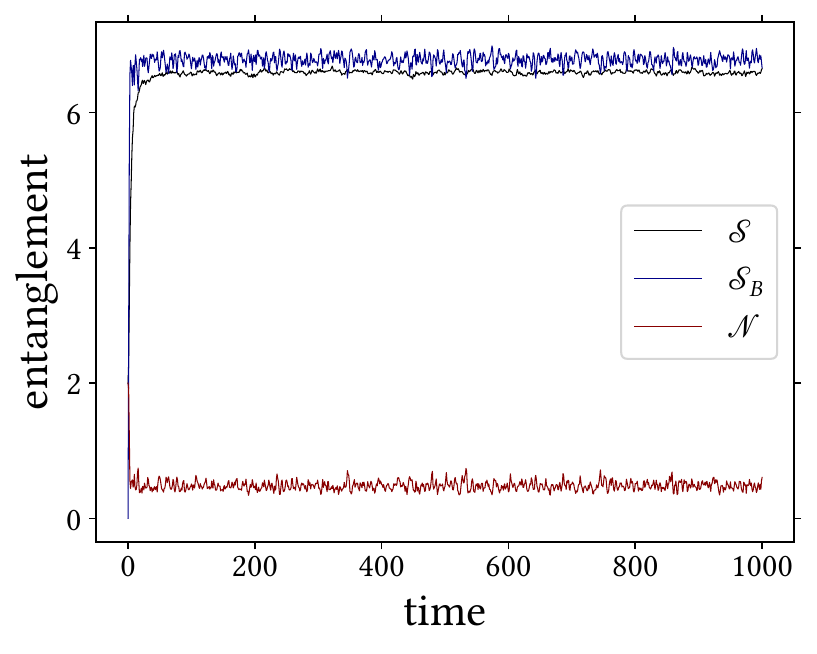}%
  \includegraphics[width=0.3\textwidth]{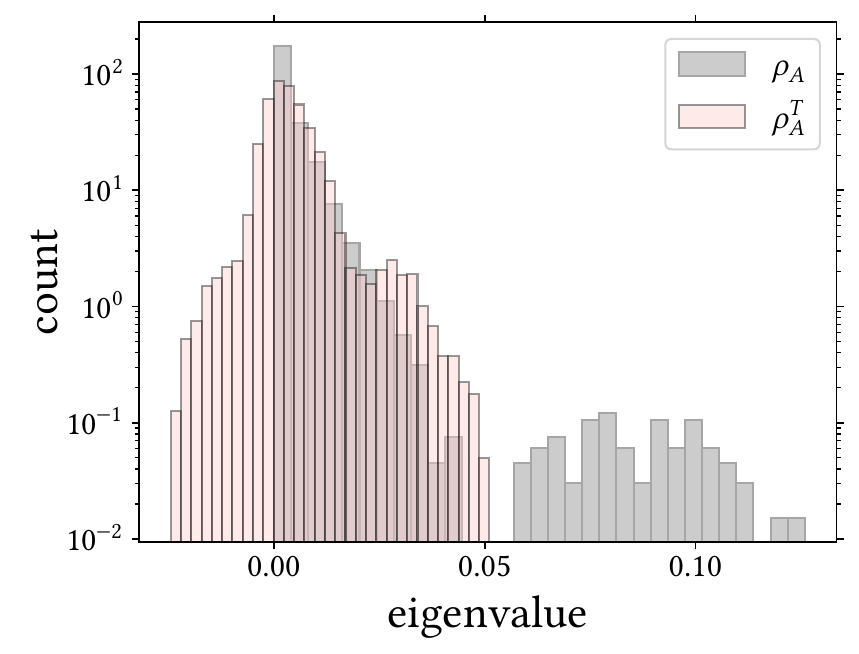}%
  \includegraphics[width=0.3\textwidth]{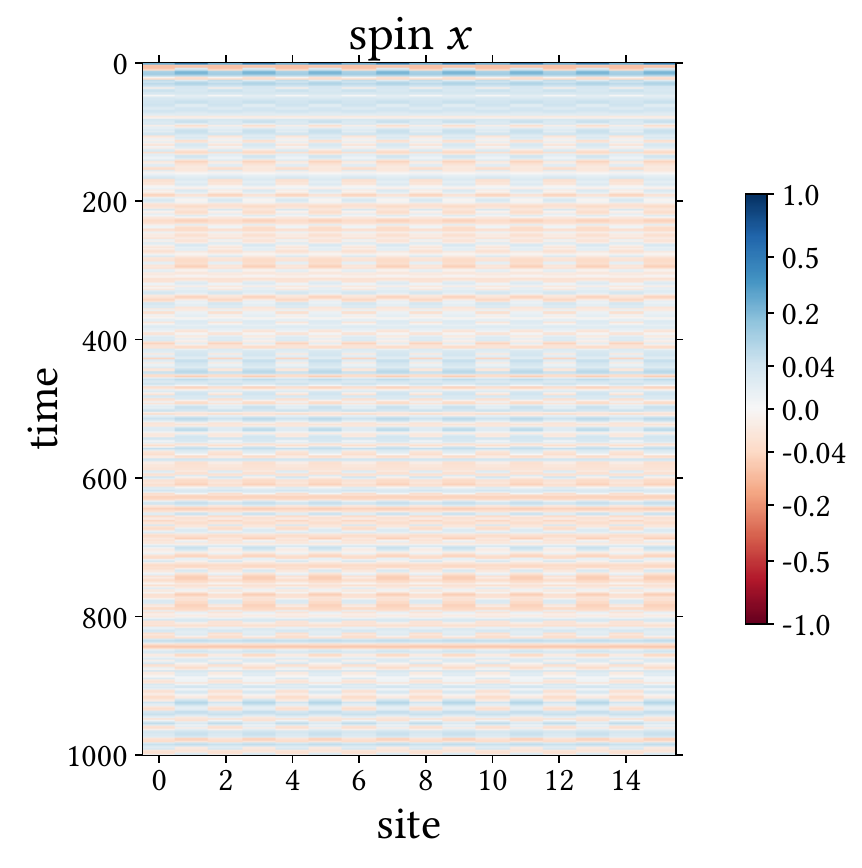}\\
  \includegraphics[width=0.3\textwidth]{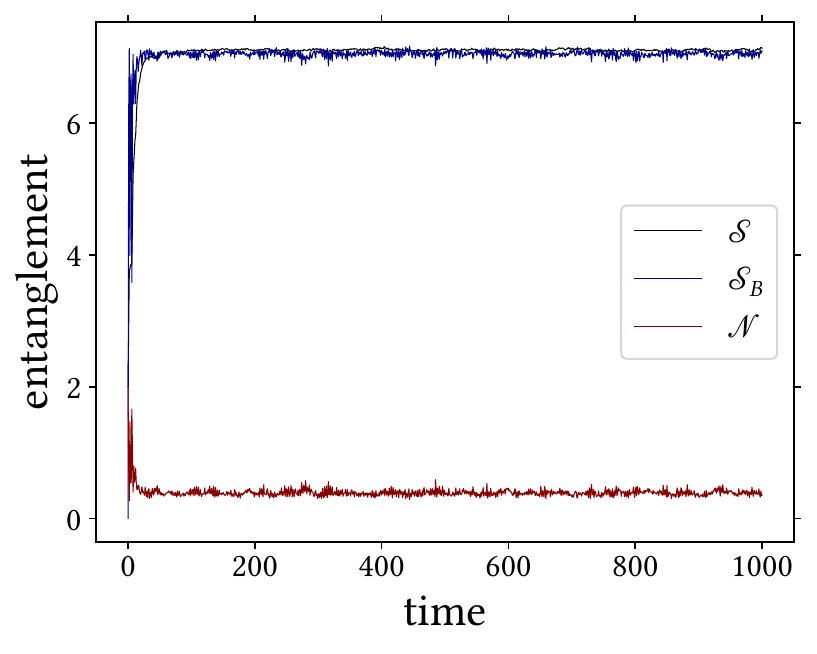}%
  \includegraphics[width=0.3\textwidth]{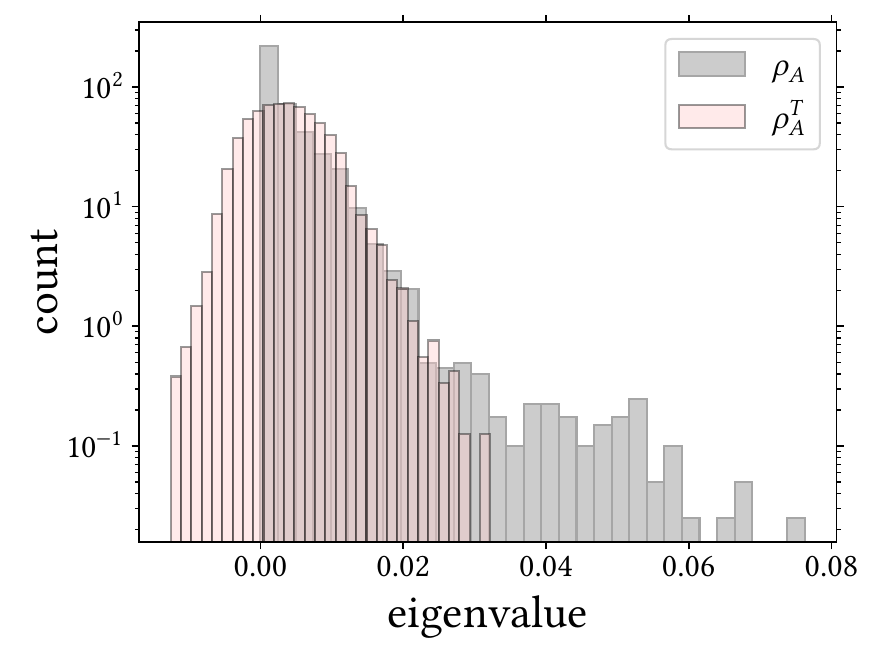}%
  \includegraphics[width=0.3\textwidth]{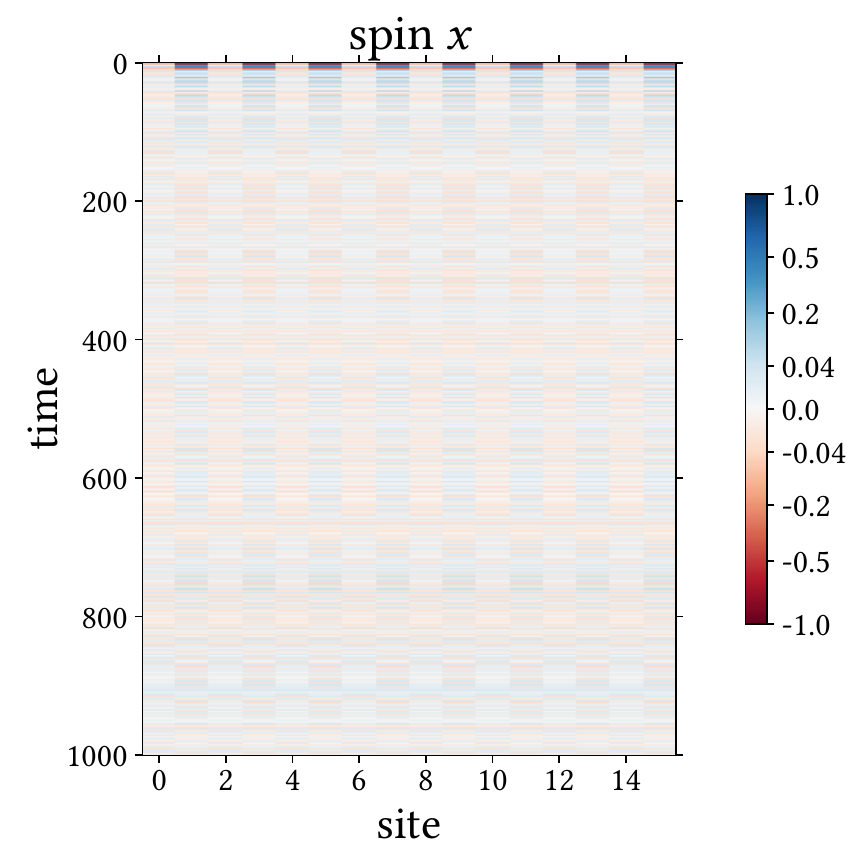}
  \caption{Entanglement and magnetization as a function of $g$ for fixed $J=0.2$. Each row corresponds to a value of $g$, with $\bar{g}=0.5, 1.0, 2.5, 5$, respectively ($2L = 16$). The first column displays the entropy $\mathcal{S}$ of $L/2$ AB cells, the environment entropy $\mathcal{S}_\textsc{b}$, and the cluster's logarithmic negativity $\mathcal{N}$. The second column contains the histograms of the cluster entanglement spectrum $\rho_\textsc{a}$ (gray) and its partial transpose $\bar{\rho}_A$ (pale red). The third column shows the mean value of $X_x$ (note the nonlinear scale  we used for the colormap to enhance the small magnetization values). The even sites correspond to the spins on sublattice A and the odd ones to sublattice B.} 
  \label{f:14e}
\end{figure*}

\section{Results}
\label{S:results}

We measure the entanglement properties of the states generated by the automaton using (i) a bipartition of the whole system $\text{S} = \text{AB} = \text{S}_1 \text{S}_2$, each partition having $|\text{S}_1| = |\text{S}_2| = L/2$ cells; (ii) a bipartition into the two sublattices A and B; and (iii) a bipartition of the cluster sublattice such that $\text{A} = \text{A}_1 \cup \text{A}_2$ with $|\text{A}_1| = |\text{A}_2| = L/4$. We define the von Neumann entanglement entropies of partition $\text{S}_1$, $\mathcal{S}$, and of partition B, $\mathcal{S}_\textsc{b}$, as
\begin{gather}
  \mathcal{S}(t) = -\Tr \rho_1(t) \log \rho_1(t), \quad  \rho_1(t) = \Tr_{\text{S}_2} \ket{\psi(t)} \bra{\psi(t)}, \\
  \mathcal{S}_\textsc{b}(t) = -\Tr \rho_\textsc{b}(t) \log \rho_\textsc{b}(t), \quad \rho_\textsc{b}(t) = \Tr_\textsc{a} \ket{\psi(t)} \bra{\psi(t)},
\end{gather}
where $\ket{\psi(t}$ is the pure state of the system AB at the $t$ iteration of the automaton \eqref{e:U}, and $\rho_1, \rho_\textsc{a}, \rho_\textsc{b}$ are the states of partitions $\text{S}_1$, A and B, respectively. The entanglement of the cluster sublattice, which is a mixed state, is conveniently characterized by the logarithmic negativity we denote $\mathcal{N}$, computed from the sum of the eigenvalues of the partial transposed state $\bar{\rho}_\textsc{a} = \rho_\textsc{a}^{T_{\textsc{a}_2}}$ \cite{Vidal-2002zr,Plenio-2005fj}:
\begin{equation}
  \label{e:logN}
  \mathcal{N}(t) = \log\Big[\sum_n |\lambda_n(t)|\Big],
\end{equation}
where
\[
 \lambda_n \in \mathrm{spec} \Big( \bar{\rho}_\textsc{a} \Big), \; \lambda = \min_n \lambda_n
\]
form the spectrum of $\bar{\rho}_\textsc{a}$ and $\lambda$ is its minimum eigenvalue (we use throughout `$\log$' for the base 2 logarithm). More detailed information on the cluster subsystem entanglement can be obtained by monitoring the entanglement spectrum $\mathrm{spec}(\rho_\textsc{a})$, the set of eigenvalues of the A subsystem density matrix \cite{Li-2008fk}, and the associated negativity spectrum $\lambda_n$, the set of eigenvalues of the partial transpose of $\rho_\textsc{a}$ \cite{Peres-1996,Znidaric-2007}, whose histograms are representative of the different phases, including topological and disordered ones. In addition to the entanglement, we follow the spatio-temporal distribution of the magnetization $\braket{\bm \sigma}$ of both subsystems, in particular to observe the eventual magnetic relaxation.

The minimum eigenvalue of the negativity spectrum, $\lambda$, is an optimal entanglement witness \cite{Znidaric-2007}; it allows us to probe the cluster subsystem's entanglement, in order to distinguish between high and low entanglement phases. The entanglement spectral properties are measured in the statistically stationary state, defined by the saturation of the half-lattice entropy $\mathcal{S}(t)$. We also introduce $m$ \cite{Znidaric-2007}, the effective number of the environment degrees of freedom coupled with the cluster subsystem $\rho_\textsc{a}$ when the whole system is in a random state. It is defined as the number of random states $\ket{r}$ (in the sense of the Haar measure) necessary to create a mixed state $R$ suitable to reproduce the same $\lambda$ as the actual state $\rho_\textsc{a}$:
\begin{equation}
  \label{e:m}
  m \;:\; R = \frac{1}{m} \sum_{n=1}^m \ket{r_n} \bra{r_n}, \quad \lambda(\rho_\textsc{a}) = \lambda(R).
\end{equation}
Note that $m$ is well defined in the random phase of the system; in this sense it cannot be compared with the mean-field order parameter. The former is a measure of the effective interacting degrees of freedom, while the latter $s_{B}$ is valid for a very large system and supposes a well defined mean value of the environment spin.

It was observed by Žnidarič et al. \cite{Znidaric-2007} that when the subsystem interacts with a large environment, its negativity eigenvalues distribution tends to become the semicircular Wigner distribution, implying a large superposition $m$ of random states in Eq.~\eqref{e:m}. With this definition, $m=1$ corresponds to a single pure random state, whose negativity spectral density is related to a convolution of two Marčenko-Pastur distributions, and $m \rightarrow \infty$ corresponds to a mixed random (thermal) state satisfying the Wigner semicircle law \cite{Bhosale-2012,Shapourian-2021}. For our model, these two cases correspond to high and low entanglement phases, respectively. Remind that the negativity of a small subsystem of a system in thermal state tends to vanish, in spite of the fact that the bipartite entropy can be maximal \cite{Hayden-2006}.

Because the evolution is unitary, the system remains in a pure state, however, due to the heterogeneity of interactions the behavior of the cluster entanglement strongly depends on the environment state. When the values of $J$ and $g$ are large (typically of order $O(1)$, since their maximal value is $\pi$), as expected the systems evolves towards a random thermal state (c.f Appendix~\ref{S:app_num} below). Therefore, we focus on values of the parameters $J,g < 1$. For this range, as predicted by the mean-field and Markov approximation, the physical properties of the cluster subsystem depend essentially on $\bar{g} = g/J$. Note that large values of $\bar{g}$ are compatible with the requirement of both $J$ and $g$ being small. Increasing the value of $\bar{g}$ we expect a transition between a topological phase similar to the one related to the cluster state, and a disordered phase, approaching the properties of a thermal state. On a first time, we use direct numerical computations to evolve the automaton \eqref{e:U}, and monitor its entanglement and magnetic properties. On a second time, we assess the topological properties in larger systems using the string order parameter \cite{Perez-Garcia-2008}.

The automaton \eqref{e:U} is initialized in a product state of the two subsystems, the entangled cluster state $\ket{C_\textsc{a}}$, defined in \eqref{e:CA}, for subsystem A, and the product state $\ket{+}^L$ for subsystem B (with magnetization in the $x$-direction, $\braket{X}=1$), and the iteration of the automaton unitary \eqref{e:U}
\begin{equation}
  \label{e:psi}
  \ket{\psi(t)} = U(J,g)^t \ket{\psi(0)}, \quad \ket{\psi(0)} = \ket{C} \otimes \ket{+}^L,
\end{equation}
gives the system state $\ket{\psi(t)}$ at step $t$. Initially, the von Neumann entropy of the system (half ladder partition) is $\mathcal{S}=2$, the environment entanglement vanishes $\mathcal{S}_\textsc{b}=0$, and the logarithmic negativity of the cluster susbsystem is $\mathcal{N} = 2$. The cluster minimum negativity eigenvalue is $\lambda = -1/4$.

\begin{figure*}[t]
  \centering
  \includegraphics[width=0.48\textwidth]{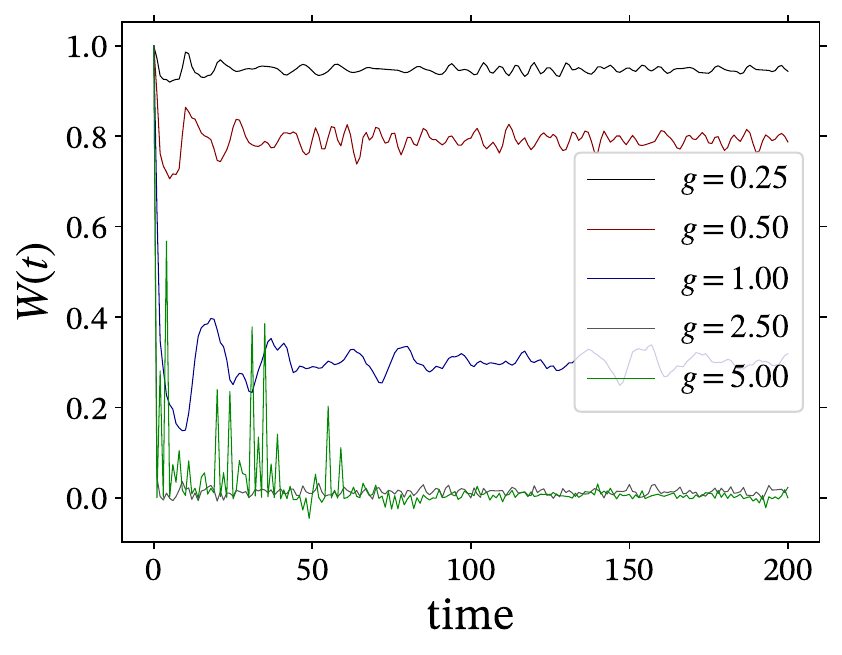}%
  \includegraphics[width=0.48\textwidth]{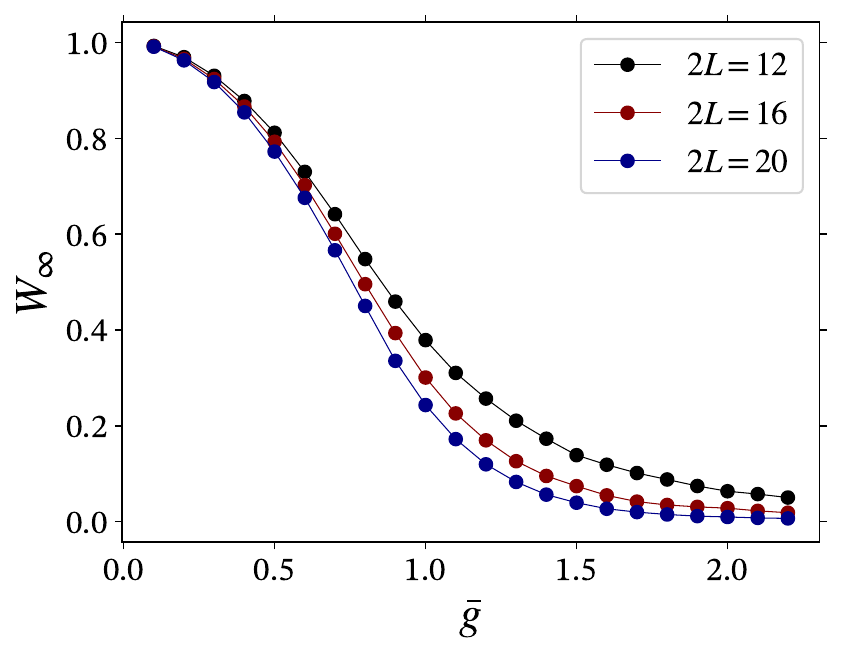}
  \caption{String order parameter $W$. (left) Time evolution for $\bar{g}=$. (right) Asymptotic value of the string order parameter $W_\infty$ as a function of $\bar{g}$, showing the transition between the topological and trivial phases for different sizes $2L=12, 16, 20$.}
  \label{f:W}
\end{figure*}

From the initial state, the system entropy start increasing up to saturation; the transient before the setting of a (statistically) stationary state strongly depends on the $(J,g)$ values. The entanglement growth results from the three and two qubits (non Clifford) gates of the automaton that efficiently scramble the information \cite{Nahum-2017qf}. In order to characterize the different regimes we studied the behavior of the minimum negativity eigenvalue $\lambda$ as a function of $J,g$. We observe in Fig.~\ref{f:lg} that for various values of the pairs $J,g$, the entanglement of the cluster subsystem as inferred from the value of $\lambda$, decreases from a maximal value when the interaction with the environment is negligible ($g$ near 0), to an unentangled state at large $g$; in fact, as we deduce from the collapse of the curves for different $J$, the minimum eigenvalue essentially depends on $\bar{g}$: we used the scaling suggested by both, the mean-field self-consistent solution \eqref{e:sb} and the effective Hamiltonian \eqref{e:lindH0}, associated to the Markov approximation. The mean-field prediction of the critical parameter $\bar{g}_c=0.425$ is in accordance with the numerical results shown in the left panel of Fig.\ref{f:lg}. Therefore, $\lambda$ distinguish between a low entanglement phase of the cluster subsystem when the interaction with the environment subsystem is strong, and a high entanglement phase for the other limit, when the environment influence is weak. In Appendix~\ref{S:app_num} we show the behavior of the negativity as a function of $\bar{g}$, displaying the entanglement transition in accordance with the behavior of $\lambda(\bar{g})$.

The right panel of Fig.~\ref{f:lg} shows the effective number of cluster-environment interacting degrees of freedom as defined by \eqref{e:m}. Around $\bar{g} \sim \bar{g}_R=2$, as suggested by the form of the effective Hamiltonian obtained in the Markov approximation \eqref{e:lindH0}, the ``size'' $m$ of the environment start to quickly grow signaling the setting of an essentially low (subsystem) entanglement regime. Below this value, the cluster subsystem can be considered slightly perturbed by the environment and one may expect the topological phase born from the initial cluster state, to be always present.

Figure~\ref{f:14e} displays the phenomenology of the cluster-free spins model through the evolution of the entanglement and the magnetization. It presents the entanglement entropies and negativity, and their corresponding spectra, computed in the stationary state, and the space-time evolution of the magnetization, for a system initially in the cluster state. We choose four different sets of parameters (rows in the figure): $J=0.2$ and $g=0.1, 0.2, 0.5, 1.0$, corresponding to $\bar{g}=0.5, 1.0, 2.5, 5$. By comparison with the results of Fig.~\ref{f:14e}, we observe that $\bar{g}=0.5, 1$ (rows 1, 2) are in the transition region, and $\bar{g}=2.5, 5$ (rows 3, 4) are in the random region. For smaller values of $\bar{g}$ (not shown in the figure) the system remains in the vicinity of the cluster state, with essentially the same histograms (only one non-vanishing four times degenerate large eigenvalue and the corresponding negativity).

We may distinguish, for each row of Fig.~\ref{f:14e}, qualitatively four different phases. The first one is characterized by a small subsystem B entropy (smaller than the negativity) and a linear growth of the total bipartite entropy; the negativity minimal eigenvalue is near its clean value and the entanglement spectrum has only one sharp pic in addition to the highly degenerated zero-eigenvalue; the spin relaxation is monotone. The second row corresponds to a state dynamically different from the first one as evidenced by the rapid saturation of the entropy and the oscillations of the relaxing magnetization; the entanglement spectrum possesses a large gap and the negativity spectrum, exhibits three enlarged pics around the cluster original eigenvalues. While in the first case (row 1), the cluster is essentially in a pure state ($m=1$), the second case (row 2) belongs to the transition region, of growing environment influence, and the cluster subsystem state is in a mixed state ($m \sim 2^8$)

For larger values of $\bar{g}$ the system sets in a random state and then relaxes towards a non-magnetic state as can be observed in the third and fourth rows of  Fig.~\ref{f:14e}. In the entanglement phase transition region, spins of the cluster and free sublattices initially align but, over longer times, alternate their orientation.  This behavior corresponds to what is expected from \eqref{e:ML}: The $-\bar{g}/2$ field created by the environment favors the $-1$ eigenvalue of $X$ (red in the figure); while in the high entangled phase, the mean-field does not reverse its sign which has been selected by the initial condition, at larger values of $\bar{g}$, in the random phases, the environment generates an adverse applied field leading to a rapid exchange between the $\pm 1$ states, of the decaying initial product $\ket{+}$ state.

Moreover, we observe that one may distinguish a phase in which the negativity spectrum is near a Marčenko-Pastur distribution (row three), and another phase closer to a thermal state with a negativity spectrum of the Wigner semicircular type. Comparing with right panel of Fig.~\ref{f:lg}, wee see that the third row is in the region $m\sim 2^7$, while the results of row four correspond to $m \sim 2^9$, meaning that the automaton approach a thermal state when the environment is large enough. Interestingly, here the control parameter is, instead of the Hilbert space dimension of the environment, the strength of the system-environment coupling.

In summary, the parameter $\bar{g}$, in accordance with the results of \S~\ref{S:model}, controls the phase transition between cluster and random phases and, within the random phases, through the effective number of the environment degrees of freedom, a transition between an entangled phase of the cluster subsystem and a thermal phase, in which the cluster's entanglement tends to vanish. 

To assess the persistence of the cluster topological phase in the presence of the decoherence effect due to the environment interaction, we measure the string order parameter \cite{Perez-Garcia-2008,Smacchia-2011,Ohta-2016} 
\begin{equation}
  \label{e:W}
  W(t,L) = (-1)^L \braket{\psi(t)| Z_1 Y_2 \Big( \prod_{x=3}^{L-2} X_x \Big) Y_{L-1} Z_L |\psi(t)},
\end{equation}
where the state $\ket{\psi(t)}$ is reached by application of the automaton $U$ for $t$ time steps, starting from the cluster state \eqref{e:psi}. For the cluster state with \emph{open} boundary conditions $W=1$, while it vanishes in a topologically trivial phase (in the absence of topological edge states). Note that the inner operator in $W$ is the parity operator associated with the $\mathbb{Z}_2 \times \mathbb{Z}_2$ symmetry \cite{denNijs-1989,Kennedy-1992}: $W$ is sensitive to the nonlocal correlation between the two edges of the spin chain, since the cluster Hamiltonian possesses, with open boundary conditions, a four times degenerated ground state \cite{Son-2012}.

Results of the numerical computation of \eqref{e:W} are shown in Fig.~\ref{f:W}. We observe that the prediction of the mean-field and the Markov approximations are confirmed by the behavior of the string order parameter: the topological phase extends to finite values of $\bar{g}$. We observe that $W(t)$ relaxes rapidly to an asymptotic mean value $W_\infty$. The topological phase at small $\bar{g}$ and the trivial phase at larger $\bar{g}$, clearly appears in the right panel of Fig.~\ref{f:W}, even if finite size effects are present. Note that for $\bar{g}=2$ we are already in the trivial phase, in accordance with the results of $m(\bar{g})$ displayed in Fig.~\ref{f:lg} (see also the third row of Fig.~\ref{f:14e}). It is also worth noting that the behavior of $W$ supports the idea that the negativity spectrum contains information about the state topology: the cluster spectrum reduces to two symmetric pics around a central degenerated zero eigenvalue, increasing $\bar{g}$ the degenerated eigenvalues spread around the original cluster eigenvalues, but the structure of the spectrum consisting in the direct sum of two blocs do not change (rows 1 and 2 of Fig.~\ref{f:14e}), and the phase extends from the cluster phase, where $W$ relaxes to a finite value (Fig.~\ref{f:W}, $\bar{g}=0.25, 0.5, 1.0$); the merging of the three pics (row 3) signals a new distinct phase (row 3), where $W$ relaxes to a vanishing value (Fig.~\ref{f:W}, $\bar{g}=2.5$, and right panel).

We conclude that the interaction with the environment do not destroy the topological order (protected by symmetry) characteristic of the cluster state. Therefore, some features of the entanglement transition in the present model are reminiscent to the entanglement transitions induced by measurement observed in monitored random circuits \cite{Fisher-2023}. This observation leads to the possibility to implement the generation of long-range entangled states using the subsystem mixed states of a unitary evolving heterogeneous system, in analogy to the protocols using measurement and classical communication \cite{Lu-2023a,Tantivasadakarn-2024}.

\section{Conclusion}
\label{S:conclusion}

We investigated how the entanglement properties of a cluster spin chain depend on its exchange coupling with a set of independent spins. By increasing the coupling strength the effective number of the environment degrees of freedom, defined as the number of random states necessary to reproduce the statistical distribution of the cluster's negativity eigenvalues, we observed a transition between the cluster topological phase, associated to the initial cluster state, to a disordered phase, associated to a random state. The random state itself undergoes an entanglement transition through which the bipartite (system and environment) entanglement entropy increases but the cluster state entanglement disappears, as can be witnessed by the minimum negativity eigenvalue. In the random phase, the subsystem high entanglement phase is consistent with a convolution of Marčenco-Patur distributions (as is our case), while the low entanglement phase shows the usual Wigner distribution, in accordance with simulations on a quantum computer \cite{Liu-2023}.

This phenomenology of entanglement transition in a unitary evolving system with heterogeneous interactions, such that one can distinguish between the cluster subsystem and the free spins subsystem, which mimics an environment, is in many respects similar to the one obtained in monitored systems. Instead of changing the empirical measurement frequency, the present model allows the existence of different phases by changing the coupling strength and keeping unitary dynamics.

It would be interesting to implement our automaton in a present day quantum computer \cite{Preskill-2023,Fauseweh-2024}. Recently quantum simulations were performed to demonstrate the measurement-induced entanglement transition \cite{Koh-2023,Hoke-2023} in noisy quantum processors, and preparation of useful entangled states using engineered dissipation through delayed measurements on ancilla qubits in a transverse field Ising model \cite{Mi-2024}. Using similar technics it should be possible to apply swap gates, as already done by Liu et al.\ \cite{Liu-2023}, and realizing the cluster three-body interaction \cite{Petiziol-2021} to implement the automaton gate $U$ of Eq.~\eqref{e:U}. This would allow to demonstrate the out-of-equilibrium entanglement transition, without recurring to measurements, monitoring the subsystem state by quantum state tomography.

\begin{acknowledgments}
  K. S. thanks D. Willsch and F. Jin for technical help. K. S. acknowledges support from the German Federal Ministry of Education and Research (BMBF), the funding program Quantum technologies -- from basic research to market, project QSolid (Grant No. 13N16149).
\end{acknowledgments} 

\begin{figure}[t]
  \centering
  \includegraphics[width=0.48\textwidth]{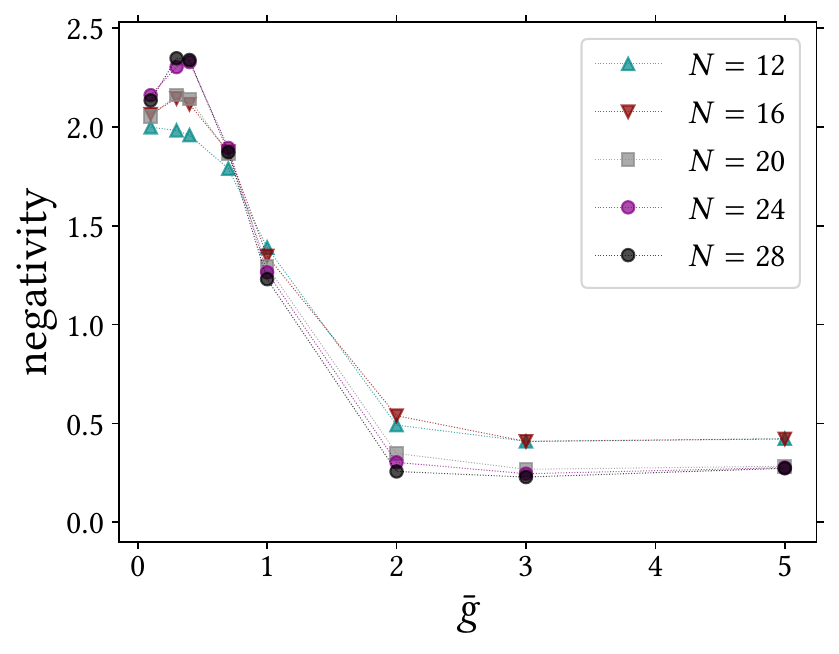}
  \caption{Logarithmic negativity as a function of $\bar{g}=0.1, 0.3, 0.4, 0.7, 1, 2, 3, 5$, for different values of the number of spins $N=12,16,20,24,28$. The two vertical lines refer to the mean-field $\bar{g}_c = 0.425$ and Markov $\bar{g}_R=2$ critical parameters.}
  \label{f:neg_mean}
\end{figure}

\appendix
\section{Self-consistent equation}
\label{S:app_sb}

To find the ground state of the mean-field Hamiltonian \eqref{e:HkMFm}, we should compute the eigenvectors of the Bogoliubov transformation, corresponding to the columns of $\T$ in Eq.~\eqref{e:T}. We find
\begin{equation}
  \frac{|d_k|}{\sqrt{|d_k|^2 + 2g^2}} \begin{pmatrix}
    \frac{\I g}{\bar{d}_k} \\ \frac{g}{\bar{d}_k} \\ 0 \\ 1
  \end{pmatrix} , \quad
  \frac{|d_k|}{\sqrt{|d_k|^2 + 2g^2}} \begin{pmatrix}
    \frac{g}{\bar{d}_k} \\ -\frac{\I g}{\bar{d}_k} \\ 1 \\ 0
  \end{pmatrix},
\end{equation}
for the two 0 eigenvalues, and
\begin{equation}
  \frac{|g|}{|\varepsilon_k|} \begin{pmatrix}
    \frac{\varepsilon_k + d_k}{2g\I} \\
    \frac{\varepsilon_k - d_k}{2g} \\
    \I \\
    1
  \end{pmatrix} , \quad
  \frac{|g|}{|\varepsilon_k|} \begin{pmatrix}
    -\frac{\varepsilon_k - d_k}{2g\I} \\
    -\frac{\varepsilon_k + d_k}{2g} \\
    \I \\
    1
  \end{pmatrix} 
\end{equation}
for the $\mp \varepsilon_k$ eigenvalues, respectively, where $d_k = 2g s_\textsc{b} - 2J s_\textsc{b}^2 \E^{2\I k}$.

The state which is annihilated by any Bogoliubov annihilation operator (for $k$ and $-k$)
\begin{equation}
  \ket{GS} = \prod_{k>0} A_k A_{-k} B_k B_{-k} \ket{0},
\end{equation}
gives us the ground state
\begin{widetext}
\begin{equation}
  \label{e:GS}
  \ket{GS} = \frac{1}{G}\prod_{k>0} \Big\{1 + \I a^\dagger_{k} a^\dagger_{-k} + \I \frac{\varepsilon_k}{g} a^\dagger_{k} b^\dagger_{-k} - \I \frac{\bar{d}_k}{g} a^\dagger_{-k} b^\dagger_{k} 
  + \I \Big( \frac{\bar{d}_k }{g} \frac{\varepsilon_k +  d_k }{2g} + 1 \Big) b^\dagger_k b^\dagger_{-k} 
  + \Big( \frac{\bar{d}_k }{g} \frac{\varepsilon_k - d_k }{2g} - 1\Big) a^\dagger_k a^\dagger_{-k} b^\dagger_k b^\dagger_{-k} \Big)
    \Big\} \ket{0},
\end{equation} 
\end{widetext}
where
\begin{equation}
  G^2 = \prod_{k>0} \frac{\varepsilon_k^2}{g^2} \frac{\varepsilon_k^2 - 2g^2}{g^2}
\end{equation}
is the normalization constant. Using $\ket{GS}$ \eqref{e:GS}, we can now solve the self-consistent equation
\begin{equation}
  s_\textsc{b} = 1 - \frac{2}{L} \sum_{k>0} \braket{GS|b^\dagger_k b_k + b^\dagger_{-k} b_{-k}|GS},
\end{equation}
which leads to Eq.~\eqref{e:sbint} in the main text.
\section{Lindblad equation}
\label{S:app_lind}

\begin{figure}[t]
  \centering
  \includegraphics[width=0.48\textwidth]{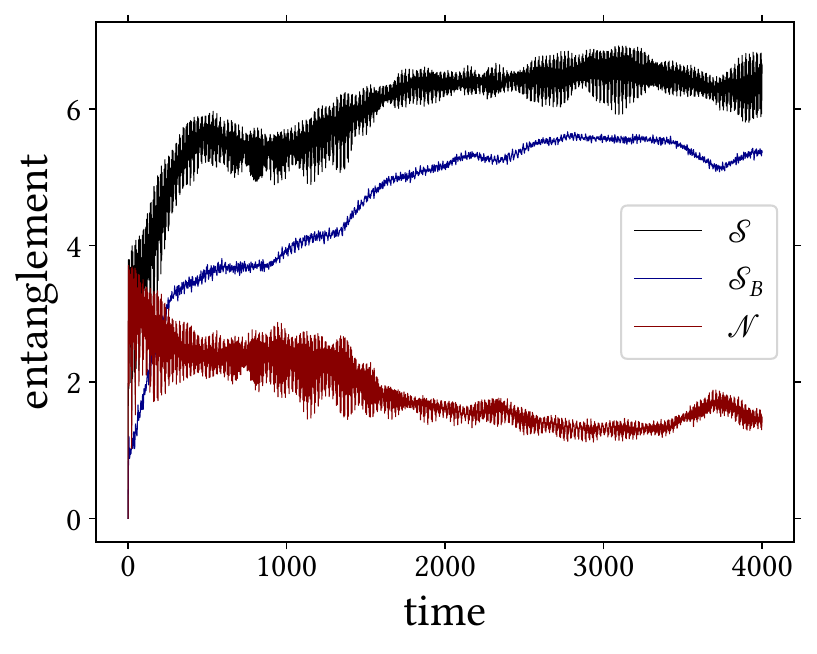}\\
  \includegraphics[width=0.48\textwidth]{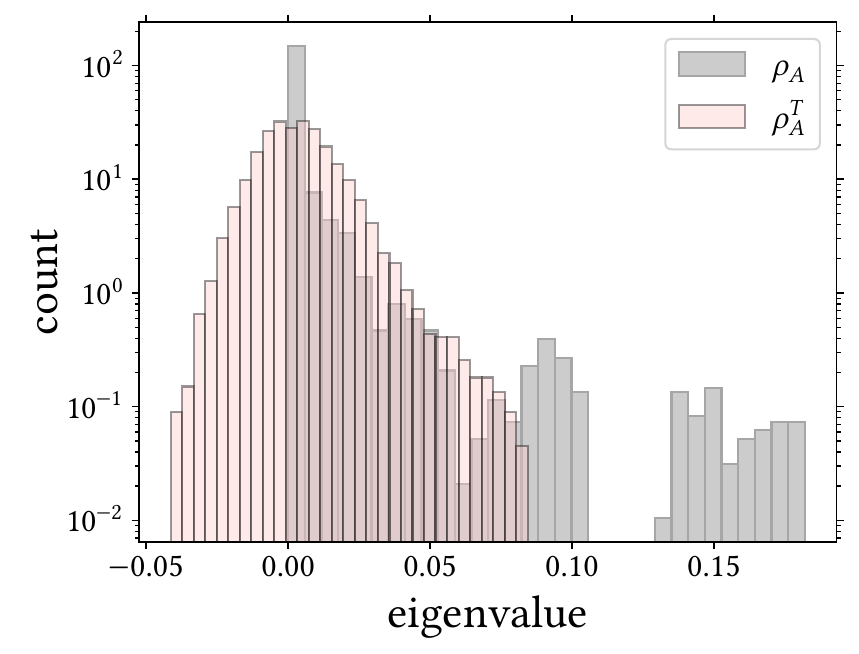}
  \caption{Entanglement evolution from an initial product state for $\bar{g}=0.33$ ($J,g=0.3,0.1)$. Entropy and negativity (top); entanglement (gray) and negativity spectra (pale red).}
  \label{f:prod}
\end{figure}

\begin{figure}[t]
  \centering
  \includegraphics[width=0.48\textwidth]{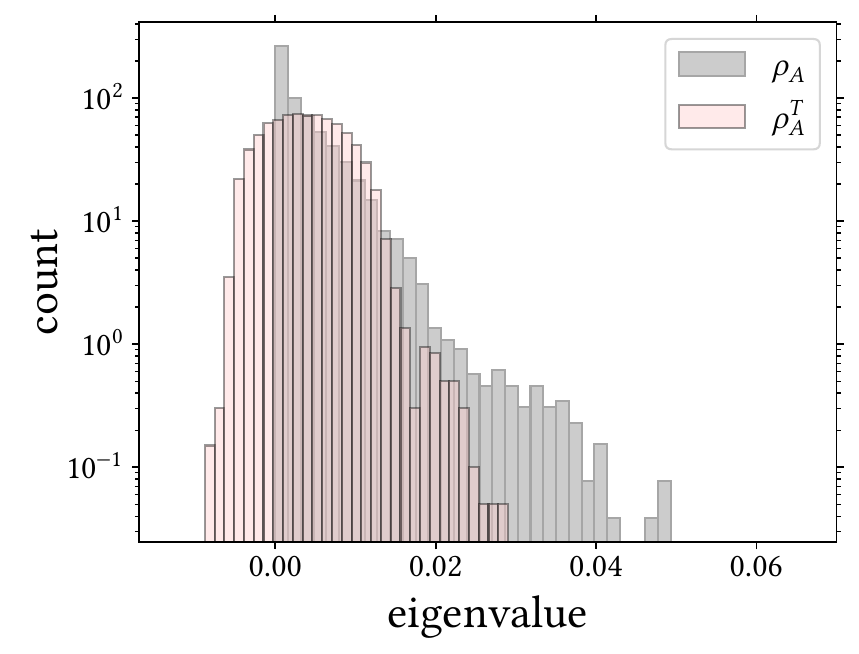}\\
  \includegraphics[width=0.48\textwidth]{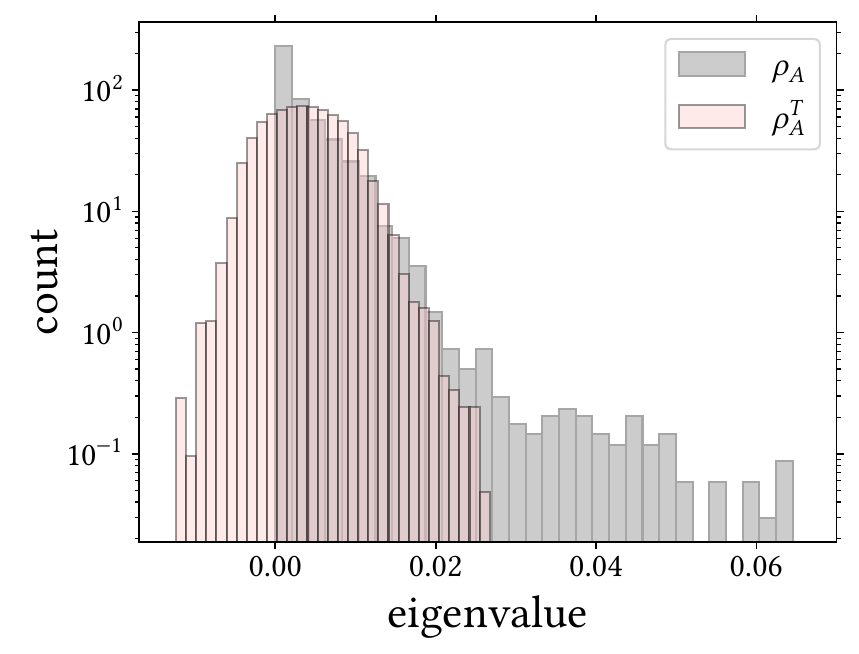}
  \caption{Entanglement and negativity spectra from initial product (top) and entangled (bottom) states for $\bar{g}=1$ and large couplings $J=1,g=1$. In both cases a random state is obtained, with a Wigner-like semicircular distribution of the negativity eigenvalues, signaling that the system approaches a thermal state.}
  \label{f:th}
\end{figure}

The usual Lindblad limit
\begin{equation}
  \label{e:lind}
  \dot{\rho}_\textsc{a}(t) = -\I [H_{\bar{0}}, \rho_\textsc{a}(t)] + \frac{\bar{g} ^2}{4} \sum_x \big[Y_x \rho_\textsc{a}(t) Y_x - \rho_\textsc{a}(t)],
\end{equation}
where
\[
  H_{\bar{0}} = -\sum_x \Big( Z_{x-1} X_x Z_{x+1} + \frac{\bar{g}}{2} X_x \Big),
\]
is obtained introducing a time step $\Delta t$ such that  $\Delta t \rightarrow 0$, and assuming that, within $\Delta t$ only one flip happens; these assumptions amount in selecting from the set of Kraus operators $M_n$, the operators $M_0$ (no flip) and $M_1$ (one flip), and making the approximation 
\begin{gather}
  \label{e:K}
  u_0 = \E^{-\I H_{\bar{0}} \Delta t} \approx 1 - \Delta t \I H_{\bar{0}} \\
  K_0 = \cos\frac{\bar{g} \sqrt{\Delta t}}{2} \approx 1 - \Delta t \frac{\bar{g} ^2}{8} \\
  K_1 = \sin\frac{\bar{g} \sqrt{\Delta t}}{2} \E^{-\I \bar{g} \Delta t X_x} Y_x  \approx \frac{\bar{g}}{2} \sqrt{\Delta t} Y_x,
\end{gather}
based on the scaling $J \rightarrow J \Delta t/\hbar$ for the unitary time evolution, and $g \rightarrow \bar{g}\sqrt{\Delta t J/\hbar}$, for the Kraus operators $K_0$ and $K_1$, corresponding to \eqref{e:sw0} and \eqref{e:sw1}, respectively  (with $\bar{g} = g/J$, and $J=\hbar=1$ units). A natural application of the Lindblad dynamics could be the case where the system is initially put in a inhomogeneous random product state, breaking the translation symmetry but preserving it statistically. In this paper we focus on homogeneous states.

\section{Complementary numerical results}
\label{S:app_num}

To assess the role of finite size effects on the entanglement transition, we investigated the behavior of the cluster subsystem negativity as a function of  $\bar{g}=g/J$ for different spin numbers $N=12,16,20,24,28$. The result is displayed in Fig.~\ref{f:neg_mean}. In spite of the small range of values of $N$ we can compute, the entanglement transition is well observed, and a tendency towards a steepening of the curve for $\bar{g} > 0.4$ with increasing $N$ is present.

When initialized in a product state $\ket{+}^L \otimes \ket{+}^L$, the system evolves towards a random phase. Remarkably, as can be seen in Fig.~\ref{f:prod}, for $\bar{g} < \bar{g}_c$, the long time dynamics is rather irregular, large fluctuations of the entanglement persist, making difficult the definition of well behaved stationary properties. For instance, the entanglement spectrum shows some stable features, such the presence of a large gap, but the distribution cannot be compared with standard distributions of simple random matrices, while the negativity spectrum is only reminiscent of the Marčenko-Pastur-like statistics, one expects for entangled near pure states \cite{Znidaric-2007,Shapourian-2021}. This contrast with the well defined stationary state of the system initialized in a cluster state, and suggests that the dynamics is not ergodic, the system evolving towards qualitatively different states in Hilbert space \cite{Verga-2023}. 

For large values of $(J,g)$ the system reach, as expected for a generic automaton, a thermal state with saturation of the global entanglement entropy and Wigner negativity spectrum of subsystems, irrespective of the initial state (Fig.~\ref{f:th}). Note that the distribution of eigenvalues is much more narrow than in the entangled phase of FIg.~\ref{f:prod}.
\bibliography{total}
\end{document}